\documentclass[showpacs,preprintnumbers,amsmath,amssymb,APSl,prd,nofootinbib,superscriptaddress,12pt]{revtex4-2}
\usepackage{graphicx}
\usepackage{subcaption}
\usepackage{xcolor}
\usepackage{hyperref}
\usepackage{amsfonts}
\usepackage{bm}
\usepackage[a4paper, margin=2cm]{geometry}
\usepackage{mathrsfs}
\usepackage{soul}
\usepackage{makecell,multirow}
\usepackage{rotating}
\usepackage[utf8]{inputenc}
\usepackage{amsmath}
\usepackage{makecell}

\usepackage{amssymb}
\usepackage{tensor}
\usepackage{graphicx}
\usepackage{bigints}
\usepackage{bbm}
\usepackage{graphicx}
\usepackage{subcaption}
\usepackage{epsf}
\usepackage{bm}
\usepackage{amsmath}
\usepackage{amsfonts}
\usepackage{amssymb}
\usepackage{graphicx}
\usepackage{tabularx}
\usepackage{multirow}
\usepackage{color}%
\providecommand{\U}[1]{\protect\rule{.1in}{.1in}}

\newcommand{\ie}{\begin{equation}}
\newcommand{\fe}{\end{equation}}

\newcommand{\mincir}{\raise
-3.truept\hbox{\rlap{\hbox{$\sim$}}\raise4.truept\hbox{$<$}\ }}
\newcommand{\magcir}{\raise
-3.truept\hbox{\rlap{\hbox{$\sim$}}\raise4.truept\hbox{$>$}\ }}

\providecommand{\U}[1]{\protect\rule{.1in}{.1in}}

\usepackage{tikz,xcolor,hyperref}

\usepackage{hyperref}             
\hypersetup{
    colorlinks=true,              
    breaklinks=true,              
    citecolor=blue,               
    linkcolor=[rgb]{0,0.5,0.9},   
    urlcolor=red,                 
    filecolor=green               
}

\definecolor{lime}{HTML}{A6CE39}
\DeclareRobustCommand{\orcidicon}{%
	\begin{tikzpicture}
	\draw[lime, fill=lime] (0,0) 
	circle [radius=0.16] 
	node[white] {{\fontfamily{qag}\selectfont \tiny ID}};
	\draw[white, fill=white] (-0.0625,0.095) 
	circle [radius=0.007];
	\end{tikzpicture}
	\hspace{-2mm}
}

\foreach \x in {A, ..., Z}{%
	\expandafter\xdef\csname orcid\x\endcsname{\noexpand\href{https://orcid.org/\csname orcidauthor\x\endcsname}{\noexpand\orcidicon}}
}

\begin{document}

\title{\Large{Particle production, absorption, scattering, and geodesics in a Schwarzschild–Hernquist black hole}}

\author{N.~Heidari\orcidA{}}
\email{heidari.n@gmail.com}

\affiliation{Center for Theoretical Physics, Khazar University, 41 Mehseti Street, Baku, AZ-1096, Azerbaijan.}
\affiliation{Departamento de Física, Universidade Federal de Campina Grande Caixa Postal 10071, 58429-900 Campina Grande, Paraíba, Brazil.}
\affiliation{School of Physics, Damghan University, Damghan, 3671641167, Iran.}


\author{A.~A.~Ara\'{u}jo~Filho\orcidB{}}
\email{dilto@fisica.ufc.br}
\affiliation{Center for Theoretical Physics, Khazar University, 41 Mehseti Street, Baku, AZ-1096, Azerbaijan.}
\affiliation{Departamento de Física, Universidade Federal da Paraíba, Caixa Postal 5008, 58051--970, João Pessoa, Paraíba,  Brazil.}
\affiliation{Departamento de Física, Universidade Federal de Campina Grande Caixa Postal 10071, 58429-900 Campina Grande, Paraíba, Brazil.}

\author{P.~H.~M.~Barros\orcidC{}}
\email{phmbarros@ufpi.edu.br}
\affiliation{Departamento de F\'{i}sica, Universidade Federal do Piau\'{i}, Teresina, Piauí, 64049-550, Brazil}
\affiliation{Departamento de Matemática e F\'{i}sica, Universidade  Estadual do Maranhão, Caxias, Maranhão, 65604-380, Brazil}



\begin{abstract}

We investigate quantum and classical signatures of a Schwarzschild black hole embedded in a Hernquist dark matter halo. Starting from the exact spherically symmetric solution describing this composite system, we analyze particle production for both bosonic and fermionic fields using semiclassical techniques. Hawking radiation is derived through Bogoliubov transformations and independently via the tunneling method with energy conservation, allowing us to identify the effective temperature, emission spectrum, and the role of dark matter parameters in suppressing particle creation. The evaporation process is examined in the high-frequency regime, leading to modified evaporation times and emission rates relative to the vacuum Schwarzschild case. We further study absorption and scattering of massless scalar waves employing a partial-wave analysis, computing phase shifts, partial and total cross sections, and assessing the impact of the Hernquist scale radius and density on these observables. Finally, null and timelike geodesics are explored to characterize light propagation and particle motion in the presence of the dark matter halo.

\end{abstract}

\maketitle

\pagebreak

\tableofcontents

\pagebreak

\section{Introduction}

The presence of dark matter has become a central element in modern astrophysics, motivated primarily by a wide range of observational anomalies that cannot be explained within luminous matter alone. Early evidence emerged from studies of the rotation curves of spiral and elliptical galaxies, which revealed mass distributions extending far beyond visible components \cite{j01}.

Subsequent analyses indicated that the dominant contribution to galactic mass budgets arises from a nonluminous component, with estimates suggesting that nearly ninety percent of the total mass of a typical galaxy may be attributed to dark matter \cite{j02}.
Within this broader astrophysical context, compact objects are not expected to exist in isolation. In particular, black holes residing at galactic centers are likely immersed in extended dark matter environments \cite{j03,j04}. This feature has motivated a growing effort to incorporate dark matter effects when modeling physical processes occurring near galactic cores \cite{j05,j06}. A variety of phenomenological density profiles have been proposed to describe such environments, each capturing distinct aspects of dark matter clustering and distribution \cite{j07,j08,j09,j10,j11,j12,j13,j14,j15,j16}. Among these models, the Dehnen family occupies a prominent role, especially in applications to dwarf and low–surface–brightness galaxies, as it allows for a continuous interpolation between different halo behaviors through suitable parameter choices \cite{j17,j18}.

In addition, attention is restricted to a specific realization of this class, namely the Dehnen–$(1,4,1)$ configuration, commonly referred to as the Hernquist profile. This choice provides a convenient and well–motivated framework for investigating the influence of a surrounding dark matter halo on black hole observables. Recent developments concerning composite systems formed by a central black hole embedded in a dark matter halo can be found in Refs.~\cite{j19,j20,j21,j22,j23,j24,j25}.
From an observational standpoint, null geodesics constitute a powerful probe of the underlying spacetime geometry. Among the various signatures derived from light propagation, the black hole shadow has attracted particular attention. It appears as a two–dimensional dark region on the observer’s sky, bounded by a luminous photon ring, and its morphology is directly governed by the spacetime structure and the parameters of the compact object. Owing to this sensitivity, shadow measurements have been extensively explored as a potential diagnostic of environmental effects, including the presence of dark matter in the vicinity of black holes \cite{j26,j27,j28,j29,j30,j31,j32,j33,j34,j35}.

Furthermore, a Schwarzschild--like black hole embedded in a Hernquist dark matter halo has been introduced recently as a new exact solution, and its thermodynamic properties, weak gravitational lensing, and constraints from Event Horizon Telescope observations were examined in Ref.~\cite{jha2025thermodynamics}. This framework has since motivated a series of follow--up investigations, including generalizations involving string clouds \cite{Ahmed:2025ttq} and electric charge \cite{Jha:2025cqf}, studies of matter accretion processes \cite{Nieto:2025apz,Shi:2025oek,Ban:2026tyh}, analyses of quasinormal spectra \cite{Feng:2025iao}, evaluations of greybody factors \cite{Lutfuoglu:2025kqp}, and further thermodynamic considerations \cite{Jumaniyozov:2025xxh}. Despite this growing body of work, several fundamental aspects remain unexplored, notably quantum particle production for both bosonic and fermionic fields, as well as absorption, scattering phenomena, and geodesic motion in this background. The present study is dedicated to addressing these open issues.

We study a black hole configuration formed by a Schwarzschild geometry coupled to a surrounding Hernquist dark matter halo and examine how this environment alters both quantum and classical processes. The analysis begins with the exact static and spherically symmetric metric of the system, which is then used to investigate particle creation for scalar and spin--$1/2$ fields within a semiclassical framework. Hawking radiation is obtained using two independent approaches: a mode analysis based on Bogoliubov coefficients and a tunneling description that incorporates energy conservation. These methods allow us to extract the radiation spectrum, the effective temperature, and the explicit dependence of the emission rate on the halo parameters. The late stages of the evaporation process are analyzed in the high--frequency limit, where the presence of dark matter modifies the mass loss rate and prolongs the black hole lifetime relative to the Schwarzschild case. In addition, we compute absorption and scattering properties of massless scalar waves through a partial--wave decomposition, determining phase shifts as well as partial and total cross sections and quantifying their sensitivity to the Hernquist scale radius and density. The study is completed with an analysis of null and timelike geodesics, which reveals how the dark matter halo affects light trajectories and particle motion around the black hole.


\section{General setup}\label{sec:BH+DM}

The interaction between black holes and their astrophysical environments has become an increasingly important topic in modern gravitational physics.  
Realistic black holes are not isolated objects: they reside in dense galactic regions influenced by baryonic matter, stellar populations, and dark matter (DM).  
Understanding how DM modifies the local spacetime geometry is essential for interpreting astrophysical observables and for accurately modelling black--hole dynamics in non-vacuum settings~\cite{liu2024probing,pezzella2025quasinormal,mollicone2025superradiance,becar2024massive,han2023tilted}.  
In this section, we construct the spacetime of a Schwarzschild black hole embedded within a Hernquist-type DM halo and derive the resulting metric.

We begin with the generalized spherical DM density family introduced by Dehnen~\cite{dehnen1993family},
\begin{equation}
\rho_{\text{s}}(r)=\rho_{\text{s}}
\left(\frac{r}{r_{\text{s}}}\right)^{-\gamma}
\left[1+\left(\frac{r}{r_{\text{s}}}\right)^{\alpha}\right]^{(\gamma-\beta)/\alpha},
\label{eq:Dehnen}
\end{equation}
where $\rho_{\text{s}}$ and ${r_\text{s}}$ denote the characteristic density and scale radius, while $(\alpha,\beta,\gamma)$ control the inner slope, asymptotic falloff, and transition sharpness of the halo.  
Specializing to the Hernquist configuration, $(\alpha,\beta,\gamma)=(1,4,1),
$, Eq.~\eqref{eq:Dehnen} reduces to the well-known form
\begin{equation}
\rho_{\text{s}}(r)=\rho_{\text{s}}\,\frac{r_{\text{s}}}{r}\left(1+\frac{r}{r_{\text{s}}}\right)^{-3}.
\label{eq:HernquistDensity}
\end{equation}
This profile is characterized by an inner cusp $\rho_{\text{s}}\propto r^{-1}$---consistent with cosmological $N$-body simulations---and a steep $r^{-4}$ asymptotic decay that ensures finite total halo mass~\cite{jha2025thermodynamics}.

Integrating Eq.~\eqref{eq:HernquistDensity} yields the enclosed DM mass,
\begin{equation}
M_{H}(r)=4\pi\int_{0}^{r}\rho_{\text{s}}(r')\,r'^{2}\mathrm{d}r'
    =2\pi\rho_{\text{s}}r_{\text{s}}^{3}\left(\frac{r}{r+r_{\text{s}}}\right)^{2},
\label{eq:MH}
\end{equation}
which behaves as $M_{H}(r)\propto r^{2}$ near the center and approaches $2\pi\rho_{\text{s}}r_{\text{s}}^{3}$ in the large-radius limit.

The gravitational field produced by the DM halo can be described by the static, spherically symmetric line element
\begin{equation}
\mathrm{d}s^{2}= -A_{1}(r)\mathrm{d}t^{2}
+ B_{1}(r)^{-1}\mathrm{d}r^{2}
+ r^{2}\mathrm{d}\Omega^{2},
\label{eq:metricHalo}
\end{equation}
with lapse function $A_{1}(r)$ and shape function $B_{1}(r)$.  
For circular geodesics, the tangential velocity $v_{t}$ satisfies
\begin{equation}
v_{t}^{2}=\frac{M_{H}(r)}{r}
= r\,\frac{\mathrm{d}}{\mathrm{d}r}\ln\!\sqrt{A_{1}(r)},
\label{eq:vtRelation}
\end{equation}
which provides a differential equation for $A_{1}(r)$.  
Integration gives
\begin{equation}
A_{1}(r)=\exp\!\left[\int\frac{2M_{H}(r)}{r^{2}}\,\mathrm{d}r\right].
\label{eq:A1general}
\end{equation}

Inserting Eq.~\eqref{eq:MH} and expanding to leading order yields  
\begin{equation}
A_{1}(r)\simeq 1-\frac{4\pi\rho_{\text{s}}r_{\text{s}}^{3}}{r+r_{\text{s}}}.
\label{eq:A1approxFinal}
\end{equation}
Following the standard assumption of isotropic halo pressure, we set
\begin{equation}
A_{1}(r)=B_{1}(r),
\end{equation}
which ensures consistency with the Einstein equations
\begin{equation}
R_{\mu\nu}-\frac12R\,g_{\mu\nu}=\kappa^{2}T^{(H)}_{\mu\nu},
\label{eq:EinsteinHalo}
\end{equation}
where $T^{(H)}_{\mu\nu}=\mathrm{diag}(-\rho_{\text{s}},p_{r},p_{t},p_{t})$.

To incorporate the Schwarzschild black hole, we extend the halo metric to
\begin{equation}
\mathrm{d}s^{2}= -f(r)\mathrm{d}t^{2}+f(r)^{-1}\mathrm{d}r^{2}+r^{2}\mathrm{d}\Omega^{2},
\label{eq:metricBHhalo}
\end{equation}
with
\begin{equation}
f(r)=A_{1}(r)+A_{2}(r).
\end{equation}
The full Einstein equations,
\begin{equation}
R_{\mu\nu}-\frac12R\,g_{\mu\nu}
=\kappa^{2}\left(T^{(H)}_{\mu\nu}+T^{(B)}_{\mu\nu}\right),
\label{eq:EinsteinTotalDM}
\end{equation}
include the Schwarzschild contribution $T^{(B)}_{\mu\nu}$, which vanishes everywhere except at the central singularity.  
Requiring that the metric reduce to the Schwarzschild form in the limit $\rho_{\text{s}}\rightarrow 0$ uniquely fixes
\begin{equation}
A_{2}(r)=-\frac{2M}{r}.
\label{eq:A2}
\end{equation}

Combining Eqs.~\eqref{eq:A1approxFinal} and~\eqref{eq:A2}, the resulting spacetime describing a Schwarzschild black hole embedded in a Hernquist dark-matter halo is
\begin{equation}
f(r)=1-\frac{2M}{r}-\frac{4\pi\rho_{\text{s}}r_{\text{s}}^{3}}{r+r_{\text{s}}}.
\label{eq:metric}
\end{equation}
This effective metric smoothly interpolates between the Schwarzschild solution at small radii and a halo-dominated gravitational potential at large distances, providing a self-consistent relativistic description of a black hole residing in a realistic galactic DM environment.




\section{Field excitations in quantum regimes}


The present section is devoted to the study of particle creation in the bumblebee black hole recently derived in this framework. Rather than beginning with general considerations, the analysis is organized according to the spin of the emitted fields. We first focus on scalar excitations and employ the semiclassical tunneling method to characterize their emission. In order to bypass the singular behavior of the metric at the horizon radius $r_{\text{h}}$, the spacetime geometry is reformulated using Painlevé--Gullstrand coordinates, which are regular across the horizon. Within this coordinate system, the dominant contribution to the tunneling probability is encoded in the imaginary part of the classical action, $\text{Im},\mathcal{S}$. This quantity is computed by isolating the pole at the horizon and evaluating the resulting contour integral via the residue method, from which the associated bosonic particle distribution $n$ is obtained. The overall strategy is aligned with the approach developed in Ref.~\cite{calmet2023quantum,AraujoFilho:2025rwr,araujo2025particleasdasd,araujo2025does,Heidari:2025oop}.

Having established the scalar case, the discussion is then redirected to fermionic modes. The emission of spin--$1/2$ particles is treated within the same tunneling perspective, but the calculation is simplified by adopting a near--horizon expansion of the relevant equations. This approximation takes into account the leading behavior responsible for particle creation and leads directly to the fermionic occupation number $n_{\psi}$. The analysis of the spinor sector follows the formalism introduced in Ref.~\cite{vanzo2011tunnelling}, suitably adapted to the present geometry.

\subsection{Perturbations of a bosonic field}


\subsubsection{Thermal radiative emission}

Before advancing to the subsequent developments, it is useful to introduce a general ansatz for the spacetime geometry under consideration. We restrict attention to static and spherically symmetric configurations, for which the metric tensor can be expressed in its most general as shown below
\begin{equation}
\mathrm{d}s^{2}= -A(r,r_{\text{s}},\rho_{\text{s}})\mathrm{d}t^{2}+\frac{1}{B(r,r_{\text{s}},\rho_{\text{s}})}\mathrm{d}r^{2}+r^{2}\mathrm{d}\Omega^{2}.
\end{equation}

The semiclassical description of black hole radiation employs scalar field modes represented by a wave function $\tilde{\Psi}$. This formulation traces back to the original treatment presented in Ref.~\cite{hawking1975particle}:
\ie
\frac{1}{\sqrt{-\Tilde{\mathrm{g}}}}\partial_{\mu}(\Tilde{\mathrm{g}}^{\mu\nu}\sqrt{-\Tilde{\mathrm{g}}} \, \partial_{\nu}\Tilde{\Psi}) = 0.
\fe
The spacetime described by the metric $\tilde{\mathrm{g}}$ was introduced at the begging of this paper in  Eq. (\ref{eq:metric}). Accordingly, the corresponding field operator takes the form
\ie
\Tilde{\Psi} = \sum_{i} \left (\Tilde{\mathrm{f}}_{i}  a_i + \bar{\Tilde{\mathrm{f}}}_{i} a^{\dagger}_{i} \right) = \sum_{i} \left( \Tilde{\mathrm{p}}_{i} b_{i} + \bar{\Tilde{\mathrm{p}}}_{i} b^{\dagger}_{i} + \Tilde{\mathrm{q}}_{i}  c_{i} + \bar{\Tilde{\mathrm{q}}}_{i}  c^{\dagger}_{i} \right ) .
\fe

The purpose of this analysis is to determine how the new parameters introduced by this new black hole, i.e., $r_{\text{s}}$ and $\rho_{\text{s}}$, modifies the mode decomposition originally employed in the standard description of black hole radiation. In the present framework, the sets of functions $\tilde{f}{i}$ and $\bar{\tilde{f}}{i}$ represent modes that are purely ingoing at the horizon, whereas $\tilde{p}{i}$ and $\bar{\tilde{p}}{i}$ correspond to modes that propagate strictly outward. A third class, given by $\tilde{q}{i}$ and $\bar{\tilde{q}}{i}$, contains solutions that do not possess any outgoing component. The operators $a_{i}$, $b_{i}$, and $c_{i}$ annihilate the corresponding modes, while their Hermitian conjugates $a_{i}^{\dagger}$, $b_{i}^{\dagger}$, and $c_{i}^{\dagger}$ generate the associated excitations. The dark matter effects are expected to deform each of these mode families, thereby altering the structure of the solutions relative to the usual Schwarzschild case.

Spherical symmetry of black holes ensures that the field equations admit a separation of variables in terms of angular harmonics. As a result, both ingoing and outgoing modes in the exterior spacetime can be expanded by them, leading to wave solutions of the form~\cite{araujo2025particleasdasd,calmet2023quantum,araujo2025does,AraujoFilho:2025rwr}:
\ie
\begin{split}
f_{\omega^\prime l m} & =  \frac{1}{\sqrt{2 \pi \omega^\prime} r }  \mathcal{F}_{\omega^\prime}(r) e^{i \omega^\prime v} Y_{lm}(\theta,\phi)\ , \\ 
p_{\omega l m} & = \frac{1}{\sqrt{2 \pi \omega} r }  \mathcal{P}_\omega(r) e^{i \omega u} Y_{lm}(\theta,\phi). 
\end{split}
\fe
For convenience, we work with null coordinates adapted to ingoing and outgoing rays, namely the advanced and retarded variables $v$ and $u$, defined for the present spacetime as
\ie
\begin{split}
& v = t + r^{*} = t  +\frac{1}{\sqrt{\left(\frac{2 M}{r}+\frac{4 \pi  \rho_{\text{s}} r_{\text{s}}^3}{r+r_{\text{s}}}-1\right)^2}} \left(\frac{2 M}{r}+\frac{4 \pi  \rho_{\text{s}} r_{\text{s}}^3}{r+r_{\text{s}}}-1\right) \\
& \times  \Bigg\{ -\frac{2 \left(2 M^2+M \left(8 \pi  \rho_{\text{s}} r_{\text{s}}^3+r_{\text{s}}\right)+2 \pi  \rho_{\text{s}} r_{\text{s}}^4 \left(4 \pi  \rho_{\text{s}} r_{\text{s}}^2-1\right)\right) \tan ^{-1}\left(\frac{-2 M+2 r-4 \pi  \rho_{\text{s}} r_{\text{s}}^3+r_{\text{s}}}{\sqrt{-4 M^2-4 M \left(4 \pi  \rho_{\text{s}} r_{\text{s}}^3+r_{\text{s}}\right)-r_{\text{s}}^2 \left(1-4 \pi  \rho_{\text{s}} r_{\text{s}}^2\right)^2}}\right)}{\sqrt{-4 M^2-4 M \left(4 \pi  \rho_{\text{s}} r_{\text{s}}^3+r_{\text{s}}\right)-r_{\text{s}}^2 \left(1-4 \pi  \rho_{\text{s}} r_{\text{s}}^2\right)^2}} \\
& -r -\left(M+2 \pi  \rho_{\text{s}} r_{\text{s}}^3\right) \ln \left|r \left(r-4 \pi  \rho_{\text{s}} r_{\text{s}}^3+r_{\text{s}}\right)-2 M (r+r_{\text{s}})\right| \Bigg\},
\end{split}
\fe
and, on the other hand, we have
\ie
\begin{split}
& v = t - r^{*} = t  -\frac{1}{\sqrt{\left(\frac{2 M}{r}+\frac{4 \pi  \rho_{\text{s}} r_{\text{s}}^3}{r+r_{\text{s}}}-1\right)^2}} \left(\frac{2 M}{r}+\frac{4 \pi  \rho_{\text{s}} r_{\text{s}}^3}{r+r_{\text{s}}}-1\right) \\
& \times  \Bigg\{ -\frac{2 \left(2 M^2+M \left(8 \pi  \rho_{\text{s}} r_{\text{s}}^3+r_{\text{s}}\right)+2 \pi  \rho_{\text{s}} r_{\text{s}}^4 \left(4 \pi  \rho_{\text{s}} r_{\text{s}}^2-1\right)\right) \tan ^{-1}\left(\frac{-2 M+2 r-4 \pi  \rho_{\text{s}} r_{\text{s}}^3+r_{\text{s}}}{\sqrt{-4 M^2-4 M \left(4 \pi  \rho_{\text{s}} r_{\text{s}}^3+r_{\text{s}}\right)-r_{\text{s}}^2 \left(1-4 \pi  \rho_{\text{s}} r_{\text{s}}^2\right)^2}}\right)}{\sqrt{-4 M^2-4 M \left(4 \pi  \rho_{\text{s}} r_{\text{s}}^3+r_{\text{s}}\right)-r_{\text{s}}^2 \left(1-4 \pi  \rho_{\text{s}} r_{\text{s}}^2\right)^2}} \\
& -r -\left(M+2 \pi  \rho_{\text{s}} r_{\text{s}}^3\right) \ln \left|r \left(r-4 \pi  \rho_{\text{s}} r_{\text{s}}^3+r_{\text{s}}\right)-2 M (r+r_{\text{s}})\right| \Bigg\}.
\end{split}
\fe
Here $r^{*}$ represents the tortoise coordinate, introduced through
\ie
\mathrm{d}r^{*} = \frac{\mathrm{d}r}{\sqrt{A(r,r_{\text{s}},\rho_{\text{s}})B(r,r_{\text{s}},\rho_{\text{s}})}}.
\fe

The influence of additional parameters on the null coordinates can be inferred by examining the trajectory of a test particle propagating in the fixed background spacetime. Parametrizing this worldline with an affine parameter $\varsigma$, the particle’s four--momentum is expressed as
\ie
p_{\mu}=\Tilde{\mathrm{g}}_{\mu\nu}\frac{\mathrm{d}x}{\mathrm{d}\varsigma}^\nu.
\fe
Notice that the momentum conservation holds along the geodesic motion, which allows one to write
\ie
\mathcal{L} = \Tilde{\mathrm{g}}_{\mu\nu} \frac{\mathrm{d}x^\mu}{\mathrm{d}\varsigma} \frac{\mathrm{d}x^\nu}{\mathrm{d}\varsigma}.
\fe
For timelike trajectories one fixes $\mathcal{L}=-1$ and chooses the affine parameter $\varsigma$ to coincide with the proper time $\tau$. The situation of interest here involves null propagation, for which $\varsigma$ remains an arbitrary affine parameter. Specializing to a static, spherically symmetric spacetime and focusing on purely radial null paths, the angular momentum is set to zero, $p_{\varphi}=L=0$, and the motion is confined to the equatorial plane, $\theta=\pi/2$. Under these conditions, the resulting relations reduce to
\ie
E =  A(r,r_{\text{s}},\rho_{\text{s}}) \dot{t}.
\fe
We define the conserved energy by setting $p_{t}=-E$, and denote derivatives along the trajectory by an overdot with respect to the affine parameter $\varsigma$. Substituting these definitions into the geodesic relations introduced above yields
\ie
\label{sdasdxx}
    \left( \frac{\mathrm{d}r}{\mathrm{d}\varsigma} \right)^2 = \frac{E^2}{A(r,r_{\text{s}},\rho_{\text{s}})B(r,r_{\text{s}},\rho_{\text{s}})^{-1}}.
\fe

A straightforward rearrangement of the preceding relations leads to
\ie
\frac{\mathrm{d}}{\mathrm{d}\varsigma}\left(t\mp r^{*}\right) = 0.
\fe
The retarded coordinate can be rewritten as
\ie
\label{scffssss}
\frac{\mathrm{d}u}{\mathrm{d}\varsigma}=\frac{2E}{A(r,r_{\text{s}},\rho_{\text{s}})}.
\fe

To construct the retarded coordinate along an ingoing null trajectory, we parametrize the geodesic by an affine parameter $\varsigma$ and treat $u$ as a derived function of this parameter. The starting point is not the coordinate $u(\varsigma)$ itself, but the radial evolution of the geodesic, $r(\varsigma)$. Once this dependence is established, the retarded coordinate follows directly from the integral relation given in Eq.~$\eqref{scffssss}$. The functional form of $u(\varsigma)$ is crucial in the analysis, as it determines the Bogoliubov coefficients governing the black hole emission spectrum.
The explicit evaluation proceeds by inserting the metric functions $A(r,r_{\text{s}},\rho_{\text{s}})$ and $B(r,r_{\text{s}},\rho_{\text{s}})$ into the square root integral appearing in Eq.~$\eqref{sdasdxx}$. This integral is computed along the ingoing null path, extending from the event horizon
$r_{\text{h}}=\frac{1}{2}\left(\sqrt{\left(-2 M-4 \pi \rho_{\text{s}} r_{\text{s}}^{3}+r_{\text{s}}\right)^{2}+8 M r_{\text{s}}}+2 M+4 \pi \rho_{\text{s}} r_{\text{s}}^{3}-r_{\text{s}}\right)$
to an arbitrary radial position $r$, corresponding to the interval $\varsigma\in[0,\varsigma]$ along the geodesic. Performing this integration yields
\ie
r = \frac{1}{2} \left(\sqrt{\left(-2 M-4 \pi  \rho_{\text{s}} r_{\text{s}}^3+r_{\text{s}}\right)^2+8 M r_{\text{s}}}+2 M+4 \pi  \rho_{\text{s}} r_{\text{s}}^3-r_{\text{s}}\right) - E\varsigma.
\fe
The result is obtained by choosing the negative root in Eq.~$\eqref{sdasdxx}$, which selects the null trajectory propagating toward the horizon.

Inserting the geodesic solution $r(\varsigma)$ into the integral renders it explicit, leading to
\ie
\begin{split}
& u(\varsigma,r_{\text{s}},\rho_{\text{s}}) = -\frac{2}{\sqrt{4 M^2+4 M \left(4 \pi  \rho_{\text{s}} r_{\text{s}}^3+r_{\text{s}}\right)+r_{\text{s}}^2 \left(1-4 \pi  \rho_{\text{s}} r_{\text{s}}^2\right)^2}} \ln \left[\frac{\varsigma}{C} \right]\\
& \times \Bigg\{  2 \pi  \rho_{\text{s}} r_{\text{s}}^3 \left(\sqrt{4 M^2+4 M \left(4 \pi  \rho_{\text{s}} r_{\text{s}}^3+r_{\text{s}}\right)+r_{\text{s}}^2 \left(1-4 \pi  \rho_{\text{s}} r_{\text{s}}^2\right)^2}+4 \pi  \rho_{\text{s}} r_{\text{s}}^3-r_{\text{s}}\right)\\
& +M \left(\sqrt{4 M^2+4 M \left(4 \pi  \rho_{\text{s}} r_{\text{s}}^3+r_{\text{s}}\right)+r_{\text{s}}^2 \left(1-4 \pi  \rho_{\text{s}} r_{\text{s}}^2\right)^2}+8 \pi  \rho_{\text{s}} r_{\text{s}}^3+r_{\text{s}}\right)+2 M^2 \Bigg\},
\end{split}
\fe
or, expand $u(\varsigma,r_{\text{s}},\rho_{\text{s}})$ up to the first order in $\rho_{\text{s}}$, we have
\ie
u(\varsigma,r_{\text{s}},\rho_{\text{s}}) \approx\, \left( -4 M -\frac{32 \rho_{\text{s}} \pi  M r_{\text{s}}^3  (M+r_{\text{s}})}{(2 M+r_{\text{s}})^2}\right) \ln \left[\frac{\varsigma}{C} \right] .
\fe

An integration constant naturally appears in the solution and will be denoted by $C$. To connect this result with the null coordinates associated with ingoing and outgoing propagation, we appeal to the geometric optics identification linking the affine parameter to the advanced null coordinate. Within this correspondence, the affine parameter is expressed as $\varsigma=(v_{0}-v)/D$, where $v_{0}$ denotes the value of the advanced coordinate at horizon crossing, corresponding to $\varsigma=0$, and $D>0$ fixes the normalization of the parametrization~\cite{calmet2023quantum}.

We next examine the outgoing sector. In this manner, the Klein--Gordon equation admits outward--moving solutions of the form
\ie
p_{\omega} =\int_0^\infty \left ( \alpha_{\omega\omega^\prime} f_{\omega^\prime} + \beta_{\omega\omega^\prime} \bar{ f}_{\omega^\prime}  \right)\mathrm{d} \omega^\prime.
\fe
The coefficients $\alpha_{\omega\omega'}$ and $\beta_{\omega\omega'}$ appearing above encode the Bogoliubov transformation between mode bases~\cite{fulling1989aspects,hollands2015quantum,parker2009quantum,wald1994quantum}
\begin{equation}
\begin{split}
    \alpha_{\omega\omega^\prime}=& -i K e^{i\omega^\prime v_0}e^{\pi \left[ 2 M +\frac{16 \rho_{\text{s}} \pi  M r_{\text{s}}^3  (M+r_{\text{s}})}{(2 M+r_{\text{s}})^2} \right]\omega} \int_{-\infty}^{0} \,\mathrm{d}x\,\Big(\frac{\omega^\prime}{\omega}\Big)^{1/2}e^{\omega^\prime x}  \times e^{i\omega\left[ 4 M +\frac{32 \rho_{\text{s}} \pi  M r_{\text{s}}^3  (M+r_{\text{s}})}{(2 M+r_{\text{s}})^2}\right]\ln\left(\frac{|x|}{CD}\right)},
    \end{split}
\end{equation}
and also
\begin{equation}
\begin{split}
    \beta_{\omega\omega'} &= i Ke^{-i\omega^\prime v_0}e^{-\pi \left[2 M +\frac{16 \rho_{\text{s}} \pi  M r_{\text{s}}^3  (M+r_{\text{s}})}{(2 M+r_{\text{s}})^2} \right]\omega}
    \int_{-\infty}^{0} \,\mathrm{d}x\,\left(\frac{\omega^\prime}{\omega}\right)^{1/2}e^{\omega^\prime x} \times e^{i\omega\left[ 4 M +\frac{32 \rho_{\text{s}} \pi  M r_{\text{s}}^3  (M+r_{\text{s}})}{(2 M+r_{\text{s}})^2}\right]\ln\left(\frac{|x|}{CD}\right)}.
    \end{split}
\end{equation}

The presence of the parameter $r_{\text{s}}$ and $\rho_{\text{s}}$ within the mode functions signals that dark matter effects reshape the ``strength'' of the particle-production process. This deformation of the spacetime background opens an additional pathway through which quantum degrees of freedom may be transmitted outward. Despite this modification of the emission amplitude, the spectrum retains its thermal form. This property can be established by computing the following quantity:
\ie
\label{bugoli}
|\alpha_{\omega\omega'}|^2 = e^{\big[ 8 \pi M +\frac{64 \pi \rho_{\text{s}} \pi  M r_{\text{s}}^3  (M+r_{\text{s}})}{(2 M+r_{\text{s}})^2} \big]\omega}|\beta_{\omega\omega'}|^2\,.
\fe
The contribution to the emitted radiation associated with frequencies around $\omega$ is obtained by restricting attention to the spectral flux within the differential window $[\omega,\omega+\mathrm{d}\omega]$~\cite{calmet2023quantum}. Carrying out this evaluation yields
\ie
\mathcal{P}(\omega, \chi)=\frac{\mathrm{d}\omega}{2\pi}\frac{1}{\left \lvert\frac{\alpha_{\omega\omega^\prime}}{\beta_{\omega\omega^\prime}}\right \rvert^2-1}\, ,
\fe
or, by substituting Eq. (\ref{bugoli}) in the above expression, we have 
\ie
\mathcal{P}(\omega, \chi)=\frac{\mathrm{d}\omega}{2\pi}\frac{1}{e^{\left[8 \pi M +\frac{64 \pi \rho_{\text{s}} \pi  M r_{\text{s}}^3  (M+r_{\text{s}})}{(2 M+r_{\text{s}})^2} \right]\omega}-1}\,.
\fe
Confronting the result with the Planck distribution shows that
\ie
\mathcal{P}(\omega, \chi)=\frac{\mathrm{d}\omega}{2\pi}\frac{1}{e^{\frac{\omega}{T}}-1}.
\fe
In this form, the expression reduces to
\ie
\label{hawggbbds}
    T = \frac{1}{8 \pi  M+\frac{64 \pi ^2 M \rho_{\text{s}} r_{\text{s}}^3 (M+r_{\text{s}})}{(2 M+r_{\text{s}})^2}} \approx \, \frac{1}{8 \pi  M}+\frac{ \left(-M r_{\text{s}}^3-r_{\text{s}}^4\right)\rho_{\text{s}}}{M (2 M+r_{\text{s}})^2}.
\fe 

An independent evaluation based on surface gravity (which will be done in the following sections) reproduces the temperature inferred from particle creation, establishing the compatibility of these two derivations.

Requiring the Hawking temperature $T$ to be well defined and positive constrains the parameter $\rho_{\mathrm{s}}$ in a manner that depends on whether one considers the full expression or its small–$\rho_{\mathrm{s}}$ approximation. For the full temperature, $T=\left[8\pi M+\dfrac{64\pi^{2}M,\rho_{\mathrm{s}}\,r_{\mathrm{s}}^{3}(M+r_{\mathrm{s}})}{(2M+r_{\mathrm{s}})^{2}}\right]^{-1}$, and assuming $M>0$ and $r_{\mathrm{s}}>0$, the temperature is always real but becomes ill defined if the denominator vanishes, which occurs at $\rho_{\mathrm{s}}=-\dfrac{(2M+r_{\mathrm{s}})^{2}}{8\pi r_{\mathrm{s}}^{3}(M+r_{\mathrm{s}})}$; imposing $T>0$ therefore requires $\rho_{\mathrm{s}}>-\dfrac{(2M+r_{\mathrm{s}})^{2}}{8\pi r_{\mathrm{s}}^{3}(M+r_{\mathrm{s}})}$. Negative values of $\rho_{\mathrm{s}}$ are thus admissible within this bound, with the temperature diverging as the lower limit is approached and decreasing monotonically for increasing $\rho_{\mathrm{s}}$, tending to zero for large positive $\rho_{\mathrm{s}}$. If instead one adopts the small–$\rho_{\mathrm{s}}$ approximation, $T \approx \dfrac{1}{8\pi M}+\dfrac{(-Mr_{\mathrm{s}}^{3}-r_{\mathrm{s}}^{4})\,\rho_{\mathrm{s}}}{M(2M+r_{\mathrm{s}})^{2}}$, the expression is finite by construction and positivity requires $\rho_{\mathrm{s}}<\dfrac{(2M+r_{\mathrm{s}})^{2}}{8\pi r_{\mathrm{s}}^{3}(M+r_{\mathrm{s}})}$; this upper bound, however, is not a genuine physical restriction but an artifact of truncating the full result, since the exact temperature remains positive for arbitrarily large positive $\rho_{\mathrm{s}}$ and only vanishes asymptotically.

The analysis shows that a black hole sourced by a dark matter configuration emits radiation characterized by an effective temperature $T$, as given in Eq.~$\eqref{hawggbbds}$. The resulting spectrum departs from a strictly ideal blackbody and instead exhibits features typical of a greybody distribution. Up to this stage, however, the emission process has been treated kinematically, without accounting for the loss of energy carried away by the radiated quanta. Since each emitted particle lowers the black hole mass, the spacetime geometry necessarily evolves during evaporation. To incorporate this backreaction consistently, the next section employs the tunneling formalism introduced by Parikh and Wilczek~\cite{011}, which provides a dynamical description of quantum tunneling with energy conservation built in.

Furthermore, imposing the extremality condition $T=0$ allows one to identify the associated remnant mass
\ie
M_{\text{rem}} = \frac{1}{2} \left(2 \pi  \rho_{\text{s}} r_{\text{s}}^3+2 \sqrt{\pi } \sqrt{\pi  \rho_{\text{s}}^2 r_{\text{s}}^6+\rho_{\text{s}} r_{\text{s}}^4}-r_{\text{s}}\right) \approx \, \pi  \rho_{\text{s}} r_{\text{s}}^3+\sqrt{\pi } \sqrt{\rho_{\text{s}}} r_{\text{s}}^2-\frac{r_{\text{s}}}{2},
\fe
up to first order in $\rho_{\text{s}}$. In addition, the quantity $M_{\text{rem}}$ will be fundamental in the black hole evaporation lifetime, which will be analyzed in the forthcoming sections.

Requiring the remnant mass $M_{\mathrm{rem}}$ to be well defined and positive imposes simultaneous constraints on the parameter $\rho_{\mathrm{s}}$. First, the reality of the square–root term in the full expression demands $\rho_{\mathrm{s}}\big(\pi \rho_{\mathrm{s}} r_{\mathrm{s}}^{2}+1\big)\ge 0$, which allows either $\rho_{\mathrm{s}}\ge 0$ or $\rho_{\mathrm{s}}\le -1/(\pi r_{\mathrm{s}}^{2})$; however, imposing $M_{\mathrm{rem}}>0$ excludes the negative branch, because even though the square root remains real for $\rho_{\mathrm{s}}\le -1/(\pi r_{\mathrm{s}}^{2})$, the total combination yields $M_{\mathrm{rem}}<0$ for any $r_{\mathrm{s}}>0$. Therefore, a physically acceptable remnant in the full computation (no approximation) requires $\rho_{\mathrm{s}}>0$ and, more sharply, $\rho_{\mathrm{s}}>1/(8\pi r_{\mathrm{s}}^{2})$, with $\rho_{\mathrm{s}}=1/(8\pi r_{\mathrm{s}}^{2})$ giving $M_{\mathrm{rem}}=0$. If instead one considers the small–$\rho_{\mathrm{s}}$ approximation $M_{\mathrm{rem}}\approx \pi \rho_{\mathrm{s}} r_{\mathrm{s}}^{3}+\sqrt{\pi}\sqrt{\rho_{\mathrm{s}}},r_{\mathrm{s}}^{2}-r_{\mathrm{s}}/2$, the square root already forces $\rho_{\mathrm{s}}\ge 0$, and the positivity condition becomes $\pi \rho_{\mathrm{s}} r_{\mathrm{s}}^{3}+\sqrt{\pi}\sqrt{\rho_{\mathrm{s}}},r_{\mathrm{s}}^{2}-r_{\mathrm{s}}/2>0$, which is equivalent (for $r_{\mathrm{s}}>0$) to $\rho_{\mathrm{s}}>\big[(\sqrt{3}-1)^{2}/(4\pi r_{\mathrm{s}}^{2})\big]=\dfrac{2-\sqrt{3}}{2\pi r_{\mathrm{s}}^{2}}$; this approximate is slightly stronger than the exact bound because truncating the full expression shifts the zero of $M_{\mathrm{rem}}$ at small $\rho_{\mathrm{s}}$.


\subsubsection{Tunneling-based analysis}

Energy conservation is enforced by abandoning a fixed-background description of the radiation and instead treating the emission as a dynamical process. In this approach, particle production is described semiclassically as barrier penetration across the horizon, with the black hole mass adjusting continuously as quanta escape. To make this picture explicit and to ensure regularity at the horizon, the spacetime is rewritten in Painlevé--Gullstrand coordinates, where the metric admits a form suitable for tracking the tunneling trajectory. In these coordinates, the line element reads
\ie
\mathrm{d}s^{2}=-A(r,r_{\text{s}},\rho_{\text{s}})\,\mathrm{d}t^{2}+2H(r,r_{\text{s}},\rho_{\text{s}})\,\mathrm{d}t\,\mathrm{d}r+\mathrm{d}r^{2}+r^{2}\mathrm{d}\Omega^{2},
\fe
in which
\ie
H(r,r_{\text{s}},\rho_{\text{s}})=\sqrt{A(r,r_{\text{s}},\rho_{\text{s}})\left(B(r,r_{\text{s}},\rho_{\text{s}})^{-1}-1\right)}.
\fe
In this parametrization, the emission probability is determined by the imaginary part of the classical action evaluated along the tunneling path, as established in semiclassical treatments of horizon crossing \cite{parikh2004energy,vanzo2011tunnelling,calmet2023quantum}.

For a particle propagating in a generic curved spacetime, the classical action can be written as follows
\ie
\mathcal{S}(r_{\text{s}},\rho_{\text{s}}) = \int p_{\mu}\,\mathrm{d}x^{\mu}.
\fe
After separating the complex contribution to the action, all components except the radial one drop out. The term associated with the time coordinate, $p_t\,\mathrm{d}t=-\omega\,\mathrm{d}t$, does not generate any imaginary contribution and can be discarded when evaluating $\mathrm{Im}\,\mathcal{S}$. As a result, the imaginary part of the action is entirely controlled by the radial sector, leading to
\ie
\text{Im}\,\mathcal{S}(r_{\text{s}},\rho_{\text{s}}) = \text{Im}\,\int_{r_i}^{r_f} \,p_r\,\mathrm{d}r=\text{Im}\,\int_{r_i}^{r_f}\int_{0}^{p_r} \,\mathrm{d}p_r'\,\mathrm{d}r.
\fe

Adopting a canonical formulation in which the black hole mass decreases as radiation is emitted, the Hamiltonian is taken to be $H=M-\omega'$. The dynamical evolution then implies that changes in the Hamiltonian are directly tied to the energy carried away by the particle. As the emitted quantum acquires energy $\omega'$ spanning the interval $0\leq\omega'\leq\omega$, the corresponding variation satisfies $\mathrm{d}H=-\,\mathrm{d}\omega'$. Substituting this relation into the action yields the tunneling term in the form
\ie
\begin{split}
\text{Im}\, \mathcal{S}(r_{\text{s}},\rho_{\text{s}}) & = \text{Im}\,\int_{r_i}^{r_f}\int_{M}^{M-\omega} \,\frac{\mathrm{d}H}{\mathrm{d}r/\mathrm{d}t}\,\mathrm{d} r = \text{Im}\,\int_{r_i}^{r_f}\,\mathrm{d}r\int_{0}^{\omega} \,-\frac{\mathrm{d}\omega'}{\mathrm{d}r/\mathrm{d}t}\,.
\end{split}
\fe
By reshuffling the order of integration and introducing a suitable change of variables, the expression can be rewritten into a more convenient form, where the tunneling contribution becomes explicit
\ie
\frac{\mathrm{d}r}{\mathrm{d}t} = -H(r,r_{\text{s}},\rho_{\text{s}})+\sqrt{A(r,r_{\text{s}},\rho_{\text{s}}) + H(r,r_{\text{s}},\rho_{\text{s}})^2} = 1-\sqrt{\frac{2 (M-\omega')}{r}+\frac{4 \pi  \rho_{\text{s}} r_{\text{s}}^3}{r+r_{\text{s}}}} , 
\fe
thereby, we get
\ie
\begin{split}
\label{ims}
& \text{Im}\, \mathcal{S} = \text{Im}\,\int_{0}^{\omega} -\mathrm{d}\omega'\int_{r_i}^{r_f}\,\frac{\mathrm{d}r}{\left( 1-\sqrt{\frac{2 (M-\omega')}{r}+\frac{4 \pi  \rho_{\text{s}} r_{\text{s}}^3}{r+r_{\text{s}}}}\right)} \\
& \approx\, \text{Im}\,\int_{0}^{\omega} -\mathrm{d}\omega'\int_{r_i}^{r_f}\, \left( \frac{1}{1-\sqrt{2} \sqrt{\frac{M-\omega^{\prime}}{r}}}+ \frac{\pi  \sqrt{2} \rho_{\text{s}} \,r \,r_{\text{s}}^3 \sqrt{\frac{M-\omega^{\prime}}{r}}}{(M-\omega^{\prime}) (r+r_{\text{s}}) \left(\sqrt{2} \sqrt{\frac{M-\omega^{\prime}}{r}}-1\right)^2} \right)\mathrm{d}r,
\end{split}
\fe
Retaining terms only up to linear order in $\rho_{\text{s}}$, the resulting expression remains tractable. Moreover, replacing the mass parameter in the metric by $M\rightarrow M-\omega'$ introduces an explicit $\omega'$ dependence into the geometry. This modification shifts the position of the horizon, which in turn produces a simple pole at the new radial location. The imaginary contribution is then obtained by encircling this pole through a counterclockwise contour integration, leading to
\begin{eqnarray}
    \text{Im}\, \mathcal{S}(r_{\text{s}},\rho_{\text{s}})  = 4 \pi  \omega  \left(M + 2 \pi  \rho_{\text{s}} r_{\text{s}}^3 \left(1-\frac{r_{\text{s}}^2}{(2 M+r_{\text{s}}) (2 M+r_{\text{s}}-2 \omega )}\right)-\frac{\omega }{2}\right)  .
\end{eqnarray}
Following the approach developed in Ref.~\cite{vanzo2011tunnelling}, the inclusion of dark matter contributions modifies the semiclassical emission process. These terms deform the tunneling probability associated with Hawking radiation, leading to a corrected decay rate that can be written as
\ie
\Gamma \sim e^{-2 \, \text{Im}\, \mathcal{S}(r_{\text{s}},\rho_{\text{s}})}=e^{- 8 \pi  \omega  \left(M + 2 \pi  \rho_{\text{s}} r_{\text{s}}^3 \left(1-\frac{r_{\text{s}}^2}{(2 M+r_{\text{s}}) (2 M+r_{\text{s}}-2 \omega )}\right)-\frac{\omega }{2}\right)} .
\fe

It is worth noting that, in the limit $\rho_{\text{s}}\!\to\!0$, the standard Schwarzschild result is recovered
\begin{equation}
    \mathcal{P}(\omega,\chi)=\frac{\mathrm{d}\omega}{2\pi}\frac{1}{e^{8 \pi  \omega  \left(M + 2 \pi  \rho_{\text{s}} r_{\text{s}}^3 \left(1-\frac{r_{\text{s}}^2}{(2 M+r_{\text{s}}) (2 M+r_{\text{s}}-2 \omega )}\right)-\frac{\omega }{2}\right)
    }-1}.
\end{equation}

Because the tunneling probability depends explicitly on the emitted frequency, the resulting radiation does not follow an exact blackbody distribution. A closer inspection shows that this departure is intrinsic to the dynamical emission process. Nevertheless, in the limit of small energies, the spectrum simplifies and approaches a Planckian behavior, with the effect of the underlying modifications absorbed into an effective Hawking temperature. The associated particle occupation number, extracted from the tunneling rate, can thus be written as
\ie
n = \frac{\Gamma}{1 - \Gamma} = \frac{1}{e^{8 \pi  \omega  \left(M + 2 \pi  \rho_{\text{s}} r_{\text{s}}^3 \left(1-\frac{r_{\text{s}}^2}{(2 M+r_{\text{s}}) (2 M+r_{\text{s}}-2 \omega )}\right)-\frac{\omega }{2}\right)} - 1}.
\fe

\begin{figure}
    \centering
\includegraphics[scale=0.6]{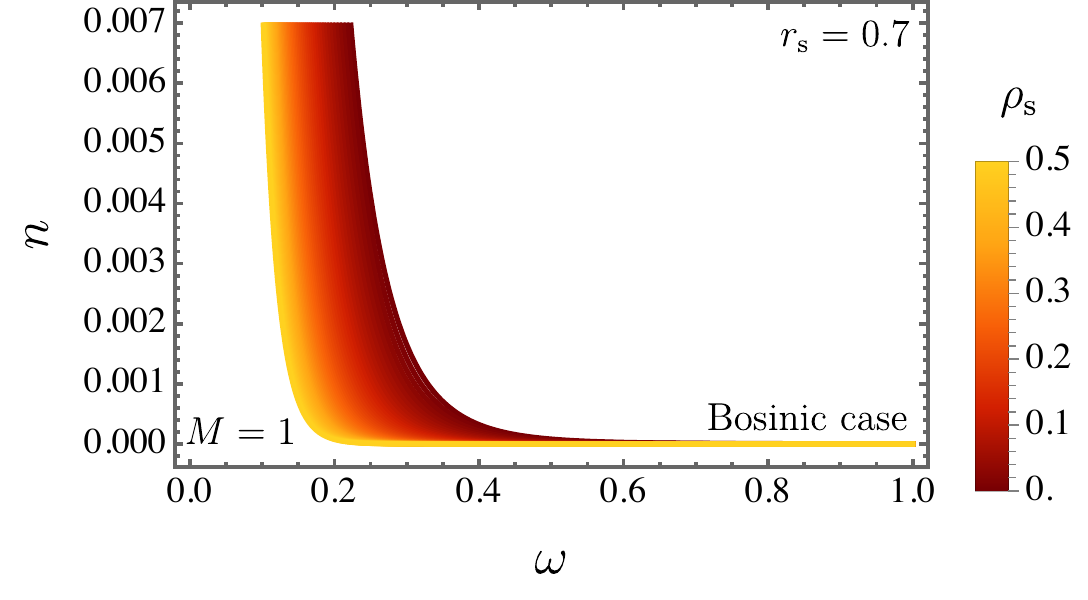}
    \caption{Frequency dependence of the bosonic number density $n$ for different choices of the parameter $\rho_{\text{s}}$, evaluated at $M=1$ and $r_{\text{s}}=0.7$.}
    \label{paerbossnn}
\end{figure}

Figure~\ref{paerbossnn} illustrates the dependence of the bosonic particle creation rate on the parameter $\rho_{\text{s}}$. As it increases, the particle yield is reduced, indicating that stronger values of this parameter hinder the production of quanta. This behavior reflects a modification of the tunneling process itself: enforcing energy conservation alters the emission amplitudes and, consequently, the power spectrum no longer coincides with the purely thermal profile associated with the Schwarzschild geometry, as we should expect.

\subsection{Fermionic modes }

Black holes emit radiation in a manner reminiscent of thermal systems, with a temperature set by their horizon properties, while the observed spectrum is shaped by transmission effects associated with the spacetime geometry. This emission process involves fields carrying different spins. Seminal work by Kerner and Mann \cite{o69}, followed by a series of refinements \cite{o75,o71,o70,o73,o74,o72}, showed that massless bosonic and fermionic excitations are characterized by the same emission temperature. Later studies extended this conclusion to vector fields, demonstrating that the inclusion of quantum corrections does not alter the Hawking temperature for spin--$1$ modes \cite{o76,o77}.

In the fermionic sector, the tunneling description is formulated in terms of the phase of the spinor field, which satisfies an equation of the Hamilton--Jacobi type. Different but equivalent implementations of this method can be found in Refs.~\cite{o84,o83,vanzo2011tunnelling}. Effects arising from the interaction between the particle spin and the spacetime connection remain regular at the horizon and manifest only as mild spin--precession corrections. These effects do not influence the leading tunneling probability and may therefore be neglected here. Moreover, emission occurs symmetrically for opposite spin states, implying that a nonrotating black hole with mass well above the Planck scale does not acquire net angular momentum through fermionic radiation \cite{vanzo2011tunnelling}.

Guided by these results, we analyze the tunneling of fermions through the near event horizon approximation of the black hole considered in this paper. The emission is evaluated using Schwarzschild--type coordinates, even though this coordinate system is singular at the horizon. Previous investigations have shown that equivalent results follow from regular coordinate choices, including generalized Painlevé--Gullstrand and Kruskal--Szekeres representations \cite{o69}. The calculation is initiated by considering a generic static and spherically symmetric line element as
\ie
\mathrm{d}s^{2} = - A(r,r_{\text{s}},\rho_{\text{s}}) \mathrm{d}t^{2} + [1/B(r,r_{\text{s}},\rho_{\text{s}})]\mathrm{d}r^{2} + C(r,\chi)[\mathrm{d}\theta^{2} + r^{2}\sin^{2}\theta \mathrm{d}\varphi^{2} ].
\fe
For a fermionic field propagating on a curved spacetime, its evolution is dictated by the covariant Dirac equation, whose structure encodes both the geometry and the spinorial nature of the particle and can be written as
\ie
\left(\Tilde{\gamma}^\mu \Tilde{\nabla}_\mu + \frac{m}{\hbar}\right) \psi(t,r,\theta,\varphi) = 0
\fe
so that we get
\ie
\Tilde{\nabla}_\mu = \partial_\mu + \frac{i}{2} {\Gamma^\alpha_{\;\mu}}^{\;\beta} \,\Tilde{\Sigma}_{\alpha\beta}
\fe and 
\ie
\Tilde{\Sigma}_{\alpha\beta} = \frac{i}{4} [\Tilde{\gamma}_\alpha,  \Tilde{\gamma}_\beta].
\fe
The spacetime--dependent gamma matrices $\tilde{\gamma}^{\mu}$ are defined so as to satisfy the local Clifford algebra relations, which are enforced through the condition
\ie
\{\Tilde{\gamma}_\alpha,\Tilde{\gamma}_\beta\} = 2 g_{\alpha\beta} \mathbb{1}.
\fe
Here, $\mathbb{1}$ denotes the identity matrix in four dimensions. Adopting this notation, a convenient explicit representation for the matrices $\tilde{\gamma}^{\mu}$ is chosen as
\begin{eqnarray*}
 \Tilde{\gamma} ^{t} &=&\frac{i}{\sqrt{A(r,r_{\text{s}},\rho_{\text{s}})}}\left( \begin{array}{cc}
\vec{1}& \vec{ 0} \\ 
\vec{ 0} & -\vec{ 1}%
\end{array}%
\right), \;\;
\Tilde{\gamma} ^{r} =\sqrt{B(r,r_{\text{s}},\rho_{\text{s}})}\left( 
\begin{array}{cc}
\vec{0} &  \vec{\sigma}_{3} \\ 
 \vec{\sigma}_{3} & \vec{0}%
\end{array}%
\right), \\
\Tilde{\gamma} ^{\theta } &=&\frac{1}{r}\left( 
\begin{array}{cc}
\vec{0} &  \vec{\sigma}_{1} \\ 
 \vec{\sigma}_{1} & \vec{0}%
\end{array}%
\right), \;\;
\Tilde{\gamma} ^{\varphi } =\frac{1}{r\sin \theta }\left( 
\begin{array}{cc}
\vec{0} &  \vec{\sigma}_{2} \\ 
 \vec{\sigma}_{2} & \vec{0}%
\end{array}%
\right).
\end{eqnarray*}%
In this expression, $\vec{\sigma}$ represents the collection of Pauli matrices, which obey the standard algebraic relations given by
\ie
 \sigma_i  \sigma_j = \vec{1} \delta_{ij} + i \varepsilon_{ijk} \sigma_k, \,\, \text{in which}\,\, i,j,k =1,2,3. 
\fe 
From this construction, the matrix $\tilde{\gamma}^{5}$ follows directly and can be written as
\begin{equation*}
\Tilde{\gamma} ^{5}=i\Tilde{\gamma} ^{t}\Tilde{\gamma} ^{r}\Tilde{\gamma} ^{\theta }\Tilde{\gamma} ^{\varphi }=i\sqrt{\frac{B(r,r_{\text{s}},\rho_{\text{s}})}{A(r,r_{\text{s}},\rho_{\text{s}})}}\frac{1}{r^{2}\sin \theta }\left( 
\begin{array}{cc}
\vec{ 0} & - \vec{ 1} \\ 
\vec{ 1} & \vec{ 0}%
\end{array}%
\right)\:.
\end{equation*}
A spinor field with polarization aligned along the outward radial direction may be represented by the following ansatz \cite{vanzo2011tunnelling}:
\begin{equation}
\psi^{+}(t,r,\theta ,\varphi ) = \left( \begin{array}{c}
\Tilde{\mathrm{H}}(t,r,\theta ,\varphi ) \\ 
0 \\ 
\Tilde{\mathrm{Y}}(t,r,\theta ,\varphi ) \\ 
0%
\end{array}%
\right) \exp \left[ \frac{i}{\hbar }\Tilde{\psi}^{+}(t,r,\theta ,\varphi )\right]\;.
\label{spinupbh} 
\end{equation} 
In the subsequent analysis, only the spin--up sector is considered, since the calculation for the polarization oriented along the inward radial direction proceeds in an analogous manner. Inserting the spinor ansatz of Eq.~(\ref{spinupbh}) into the Dirac equation on the curved background yields the following coupled equations:
\ie
\begin{split}
-\left( \frac{i \,\Tilde{\mathrm{H}}}{\sqrt{A(r,r_{\text{s}},\rho_{\text{s}})}}\,\partial _{t} \Tilde{\psi}_{+} + \Tilde{\mathrm{Y}} \sqrt{B(r,r_{\text{s}},\rho_{\text{s}})} \,\partial_{r} \Tilde{\psi}_{+}\right) + \Tilde{\mathrm{H}} m &=0, \\
-\frac{\Tilde{\mathrm{Y}}}{r}\left( \partial _{\theta }\Tilde{\psi}_{+} +\frac{i}{\sin \theta } \, \partial _{\varphi }\Tilde{\psi}_{+}\right) &= 0, \\
\left( \frac{i \,\Tilde{\mathrm{Y}}}{\sqrt{A(r,r_{\text{s}},\rho_{\text{s}})}}\,\partial _{t}\Tilde{\psi}_{+} - \Tilde{\mathrm{H}} \sqrt{B(r,r_{\text{s}},\rho_{\text{s}})}\,\partial_{r}\Tilde{\psi}_{+}\right) + \Tilde{\mathrm{Y}} m & = 0, \\
-\frac{\Tilde{\mathrm{H}}}{r}\left(\partial _{\theta }\Tilde{\psi}_{+} + \frac{i}{\sin \theta }\,\partial _{\varphi }\Tilde{\psi}_{+}\right) &= 0,
\end{split}
\fe%

Keeping only the leading semiclassical term in the expansion in $\hbar$, the fermionic phase is assumed to admit a separable structure. Within this approximation, the action splits naturally into temporal, radial, and angular pieces and can be expressed as
$\tilde{\psi}_{+} -\,\omega\, t + \Xi(r) + L(\theta,\varphi)$,
which is the form employed in the tunneling formulation of Ref.~\cite{vanzo2011tunnelling}
\begin{eqnarray}
\left( \frac{i\, \omega\, \Tilde{\mathrm{H}}}{\sqrt{A(r,r_{\text{s}},\rho_{\text{s}})}} - \Tilde{\mathrm{Y}} \sqrt{B(r,r_{\text{s}},\rho_{\text{s}})}\,  \Tilde{\Xi}^{\prime }(r)\right) +m\, \Tilde{\mathrm{H}} &=&0,
\label{bhspin5} \\
-\frac{\Tilde{\mathrm{H}}}{r}\left( L_{\theta }+\frac{i}{\sin \theta }L_{\varphi }\right) &=&0,
\label{bhspin6} \\
-\left( \frac{i\,\omega\, \Tilde{\mathrm{Y}}}{\sqrt{A(r,r_{\text{s}},\rho_{\text{s}})}} + \Tilde{\mathrm{H}}\sqrt{B(r,r_{\text{s}},\rho_{\text{s}})}\,  \Tilde{\Xi}^{\prime }(r)\right) +\Tilde{\mathrm{Y}}\,m &=&0,
\label{bhspin7} \\
-\frac{\Tilde{\mathrm{H}}}{r}\left( L_{\theta } + \frac{i}{\sin \theta }L_{\varphi }\right) &=& 0.
\label{bhspin8}
\end{eqnarray}
The precise functional form of the angular coefficients $\tilde{\mathrm{H}}$ and $\tilde{\mathrm{Y}}$ does not enter the restriction implied by Eqs.~(\ref{bhspin6}) and (\ref{bhspin8}). Instead, these equations reduce the angular dependence to the condition
$L_{\theta}+ i\,(\sin\theta)^{-1} L_{\varphi}=0 ,$ which enforces a complex structure for the function $L(\theta,\varphi)$. This requirement is common to both ingoing and outgoing fermionic configurations. When the corresponding tunneling probabilities are compared, every factor involving $L$ drops out, showing that the angular sector leaves the emission rate unchanged. For this reason, the angular contribution can be safely excluded from the remainder of the calculation.

In the case of a massless fermion, Eqs.~(\ref{bhspin5}) and (\ref{bhspin7}) give rise to two distinct solution branches:
\ie
\Tilde{\mathrm{H}} = -i \Tilde{\mathrm{Y}}, \qquad \Tilde{\Xi}^{\prime }(r) = \Tilde{\Xi}_{\text{out}}' = \frac{\omega}{\sqrt{A(r,r_{\text{s}},\rho_{\text{s}})B(r,r_{\text{s}},\rho_{\text{s}})}},
\fe
\ie
\Tilde{\mathrm{H}} = i \Tilde{\mathrm{Y}}, \qquad \Tilde{\Xi}^{\prime }(r) = \Tilde{\Xi}_{\text{in}}' = - \frac{\omega}{\sqrt{A(r,r_{\text{s}},\rho_{\text{s}})B(r,r_{\text{s}},\rho_{\text{s}})}}.
\fe

\begin{figure}
    \centering
\includegraphics[scale=0.6]{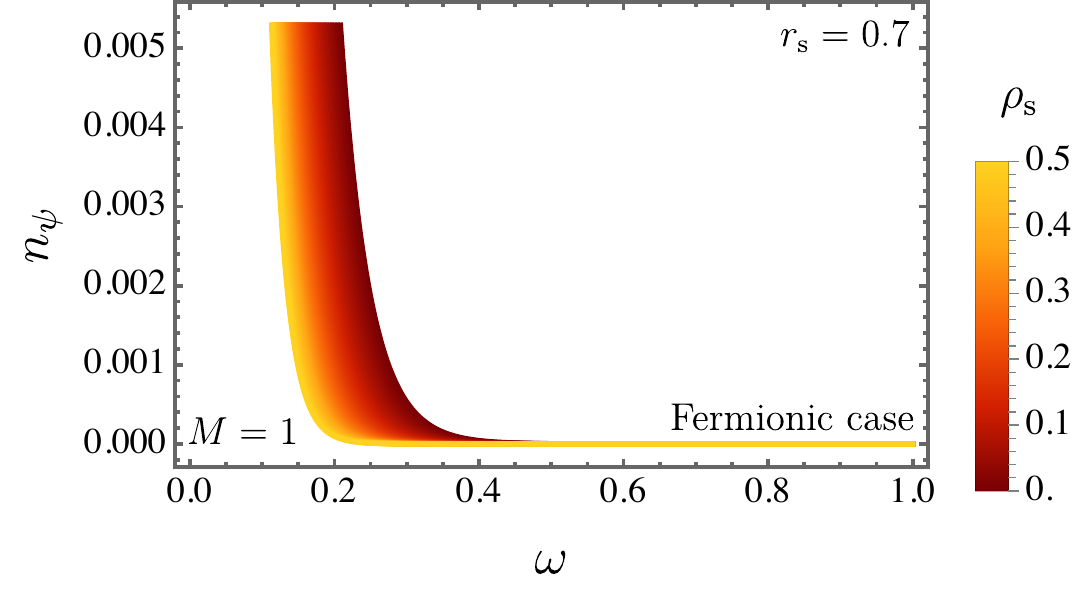}
    \caption{Frequency dependence of the fermionic number density $n_{\psi}$ for several choices of the parameter $\rho_{\text{s}}$, evaluated at $M=1$ and $r_{\text{s}}=0.7$.}
    \label{nfer}
\end{figure}

In this notation, $\tilde{\Xi}_{\text{out}}$ and $\tilde{\Xi}_{\text{in}}$ label the radial actions associated with outgoing and ingoing fermionic trajectories, respectively \cite{vanzo2011tunnelling}. The emission probability is then controlled by the difference between the imaginary components of these two solutions,
$$
\Gamma_{\psi} \propto \exp\!\left[-2\,\mathrm{Im}\!\left(\Tilde{\Xi}_{\text{out}} - \Tilde{\Xi}_{\text{in}}\right)\right].
$$
In this manner, we can obtain
\ie
 \Tilde{\Xi}_{\text{out}}(r)= -  \Tilde{\Xi}_{ \text{in}} (r) = \int \mathrm{d} r \,\frac{\omega}{\sqrt{A(r,r_{\text{s}},\rho_{\text{s}})B(r,r_{\text{s}},\rho_{\text{s}})}}\,.
\fe
The fundamental point is that, once the Einstein equations are imposed alongside the dominant energy condition, the metric functions $A(r,r_{\text{s}},\rho_{\text{s}})$ and $B(r,r_{\text{s}},\rho_{\text{s}})$ are forced to share the same zero. As a result, near the horizon location $r=r_h$, each function admits a linear expansion in the radial coordinate, of the form
\ie
A(r,r_{\text{s}},\rho_{\text{s}})B(r,r_{\text{s}},\rho_{\text{s}}) = A'(r_{\text{h}},\chi) B'(r_{\text{h}}, \chi)(r - r_{\text{h}})^2 + \dots \, .
\fe
These near horizon expansions reveal the appearance of a first order pole with a well--defined residue. Applying the standard Feynman prescription to treat this singular behavior then leads to the result:
\ie
2\mbox{ Im}\;\left(  \Tilde{\Xi}_{ \text{out}} -  \Tilde{\Xi}_{ \text{in}} \right) = \mbox{Im}\int \mathrm{d} r \,\frac{4\omega}{\sqrt{A(r,r_{\text{s}},\rho_{\text{s}})B(r,r_{\text{s}},\rho_{\text{s}})}}=\frac{2\pi\omega}{\kappa},
\fe
with the parameter $\kappa$ representing the surface gravity associated with the horizon and is defined by
\ie
\kappa = \frac{1}{2} \sqrt{A'(r_{\text{h}},\chi) B'(r_{\text{h}},\chi)} .
\fe 
Since the tunneling rate takes the form $\Gamma_{\psi}\sim e^{-\,2\pi\omega/\kappa}$, one can therefore extract the corresponding result by direct comparison
\ie
n_{\psi} = \frac{\Gamma_{\psi}}{1+\Gamma_{\psi}}  = \frac{1}{\exp \left(\frac{2 \pi  \omega }{\frac{2 \pi  \rho_{\text{s}} r_{\text{s}}^3}{\left(M \left(\frac{8 \pi  \rho_{\text{s}} r_{\text{s}}^3}{2 M+r_{\text{s}}}+2\right)+r_{\text{s}}\right)^2}+\frac{1}{M \left(\frac{8 \pi  \rho_{\text{s}} r_{\text{s}}^3}{2 M+r_{\text{s}}}+2\right)^2}}\right)+1}.
\fe
Figure~\ref{nfer} displays the dependence of the fermionic number density $n_{\psi}$ on the parameter $\rho_{\text{s}}$. As it increases, the production of fermionic quanta is progressively reduced, reproducing the same suppressive behavior observed earlier for bosonic emission.

\section{Evaporation and emission rate}

This section is devoted to a qualitative examination of a particular stage of the evaporation dynamics, namely the high--frequency regime. In this limit, the emission process is described using the \textit{Stefan--Boltzmann} law \cite{ong2018effective}
\ie
\label{sflaw}
\frac{\mathrm{d}M}{\mathrm{d}t}  =  - \alpha a \sigma T^{4},
\fe
with $\sigma$ denoting the effective emission area, $a$ is the radiation constant, $\alpha$ encodes the greybody corrections, and $T$ stands for the Hawking temperature. In the high--frequency regime, the description simplifies substantially. The absorption cross section saturates to its asymptotic value, $\sigma \to \sigma_{\mathrm{lim}} \simeq \pi \mathcal{R}^{2}$, with $\mathcal{R}$ identified as the black hole shadow radius. At the same time, the greybody factors approach unity and no longer distort the spectrum \cite{liang2025einstein}.

It is also important to recall that the dominant contribution to the emitted flux originates from effectively massless species, such as photons and neutrinos \cite{hiscock1990evolution,page1976particle}. Within the present high--frequency approximation, the shadow radius provides the relevant geometric scale governing the limiting cross section, while the remaining characteristics of the black hole spacetime enter only through this quantity and related geometric parameters
\ie
\begin{split}
& \mathcal{R} \approx \,  3 \sqrt{3} M+ \frac{18 \pi  \sqrt{3} M \rho_{\text{s}} r_{\text{s}}^3}{3 M+{\text{s}}}, \quad \sigma_{lim} \approx  \, \frac{27 M^2 \left(3 M+6 \pi  \rho_{\text{s}} r_{\text{s}}^3+r_{\text{s}}\right)^2}{(3 M+r_{\text{s}})^2} \\
& \text{and} \quad T \approx \, \, \frac{1}{8 \pi  M}+\frac{ \left(-M r_{\text{s}}^3-r_{\text{s}}^4\right)\rho_{\text{s}}}{M (2 M+r_{\text{s}})^2}.
\nonumber
\end{split}
\fe
In the present analysis, the radiation constant is further set to $a=1$. Under this assumption, the expression simplifies and can be written as
\ie
\frac{\mathrm{d}M}{\mathrm{d}t} = -\frac{27 \pi  \left(3 M+12 \pi  \rho_{\text{s}} r_{\text{s}}^3+r_{\text{s}}\right) \left(\frac{1}{\pi }-\frac{8 \rho_{\text{s}} r_{\text{s}}^3 (M+r_{\text{s}})}{(2 M+r_{\text{s}})^2}\right)^4}{4096 M^2 (3 M+r_{\text{s}})}\,.
\fe

The next step consists of explicitly evaluating the integral that follows
\ie
\begin{split}
& \int_{0}^{t_{\text{evap}}} \xi \mathrm{d}\tau = 
	- \int_{M_{i}}^{M_{f}} \mathrm{d}M
\left[ \frac{27 \pi  \left(3 M+12 \pi  \rho_{\text{s}} r_{\text{s}}^3+r_{\text{s}}\right) \left(\frac{1}{\pi }-\frac{8 \rho_{\text{s}} r_{\text{s}}^3 (M+r_{\text{s}})}{(2 M+r_{\text{s}})^2}\right)^4}{4096 M^2 (3 M+r_{\text{s}})}  \right]^{-1} .
\end{split}
\fe
Here, $t_{\text{evap}}$ represents the total evaporation time of the black hole and can thus be written in the form
\ie
\begin{split}
& t_{\text{evap}}  \approx\, \frac{4096}{243} \pi ^3 \Bigg(-3 M_{f}^3+2 \pi  \rho_{\text{s}} r_{\text{s}}^5 (9 \ln (2 M_{f}+r_{\text{s}})+2 \ln (3 M_{f}+r_{\text{s}})-9 \ln (2 M_{i}+r_{\text{s}}) \\
& -2 \ln (3 M_{i}+r_{\text{s}})) +3 M_{i}^3\\
& -\frac{6 \pi  \rho_{\text{s}} r_{\text{s}}^3 (M_{f}-M_{i}) \left(7 r_{\text{s}}^2 (M_{f}+M_{i})+2 r_{\text{s}} (3 M_{f}+M_{i}) (M_{f}+3 M_{i})+12 M_{f} M_{i} (M_{f}+M_{i})+8 r_{\text{s}}^3\right)}{(2 M_{f}+r_{\text{s}}) (2 M_{i}+r_{\text{s}})} \Bigg).
\end{split}
\fe

Requiring the Hawking temperature to vanish, $T\to0$, singles out the mass corresponding to the endpoint of the evaporation process. This condition determines a null remnant characterized by
$M_{\mathrm{rem}}\approx \pi\,\rho_{\text{s}} r_{\text{s}}^{3}+\sqrt{\pi}\,\sqrt{\rho_{\text{s}}}\, r_{\text{s}}^{2}-\tfrac{1}{2}r_{\text{s}}$.
The evolution therefore drives the black hole toward full evaporation, with the mass asymptotically approaching this bound, $M_{f}\to M_{\mathrm{rem}}$. In this regime, the total evaporation time is consequently given by
\ie
\begin{split}
	t_{\text{evap-final}} = &  \, \frac{4096 \pi ^3 M_{i}^3}{81} + \frac{16384 \pi ^4 M_{i}^3 \rho_{\text{s}} r_{\text{s}}^3}{27 (2 M_{i}+r_{\text{s}})}+\frac{57344 \pi ^4 M_{i}^2 \rho_{\text{s}} r_{\text{s}}^4}{81 (2 M_{i}+r_{\text{s}})}-\frac{31744 \pi ^4 \rho_{\text{s}} r_{\text{s}}^6}{81 (2 M_{i}+r_{\text{s}})}\\
    & +\frac{16384 \pi ^4 \rho_{\text{s}} r_{\text{s}}^6 \ln \left(-\frac{r_{\text{s}}}{2}\right)}{243 (2 M_{i}+r_{\text{s}})}-\frac{8192 \pi ^4 \rho_{\text{s}} r_{\text{s}}^6 \ln (2 M_{i}+r_{\text{s}})}{27 (2 M_{i}+r_{\text{s}})}-\frac{16384 \pi ^4 \rho_{\text{s}} r_{\text{s}}^6 \ln (3 M_{i}+r_{\text{s}})}{243 (2 M_{i}+r_{\text{s}})}\\
    & +\frac{2048 \pi ^4 M_{i} \rho_{\text{s}} r_{\text{s}}^5}{81 (2 M_{i}+r_{\text{s}})}+\frac{32768 \pi ^4 M_{i} \rho_{\text{s}} r_{\text{s}}^5 \ln \left(-\frac{r_{\text{s}}}{2}\right)}{243 (2 M_{i}+r_{\text{s}})}-\frac{16384 \pi ^4 M_{i} \rho_{\text{s}} r_{\text{s}}^5 \ln (2 M_{i}+r_{\text{s}})}{27 (2 M_{i}+r_{\text{s}})}\\
    & -\frac{32768 \pi ^4 M_{i} \rho_{\text{s}} r_{\text{s}}^5 \ln (3 M_{i}+r_{\text{s}})}{243 (2 M_{i}+r_{\text{s}})}+\frac{8192 \pi ^4 \rho_{\text{s}} r_{\text{s}}^6 \ln \left(2 \sqrt{\pi } \sqrt{\rho_{\text{s}}} r_{\text{s}}^2\right)}{27 (2 M_{i}+r_{\text{s}})}\\
    & +\frac{16384 \pi ^4 M_{i} \rho_{\text{s}} r_{\text{s}}^5 \ln \left(2 \sqrt{\pi } \sqrt{\rho_{\text{s}}} r_{\text{s}}^2\right)}{27 (2 M_{i}+r_{\text{s}})}+\frac{1024}{9} \pi ^{7/2} \sqrt{\rho_{\text{s}}} r_{\text{s}}^4+\frac{512 \pi ^3 r_{\text{s}}^3}{81}.
    \end{split}
\fe

The leading term in the result coincides with the evaporation time obtained for a Schwarzschild black hole, while the remaining contributions encode the effects associated with the dark matter parameters introduced in the model. Across the entire range explored, the final evaporation time $t_{\text{evap-final}}$ exceeds the Schwarzschild limit corresponding to $\rho_{\text{s}}=0$. This behavior indicates that, in comparison with the standard case, the presence of a nonzero $\rho_{\text{s}}$ slows down the mass depletion, so that larger values of this parameter systematically prolong the black hole lifetime. This result will be corroborated in the subsequent analysis concerning emission rate as well.

\begin{figure}[ht]
    \centering
    \includegraphics[width=82mm]{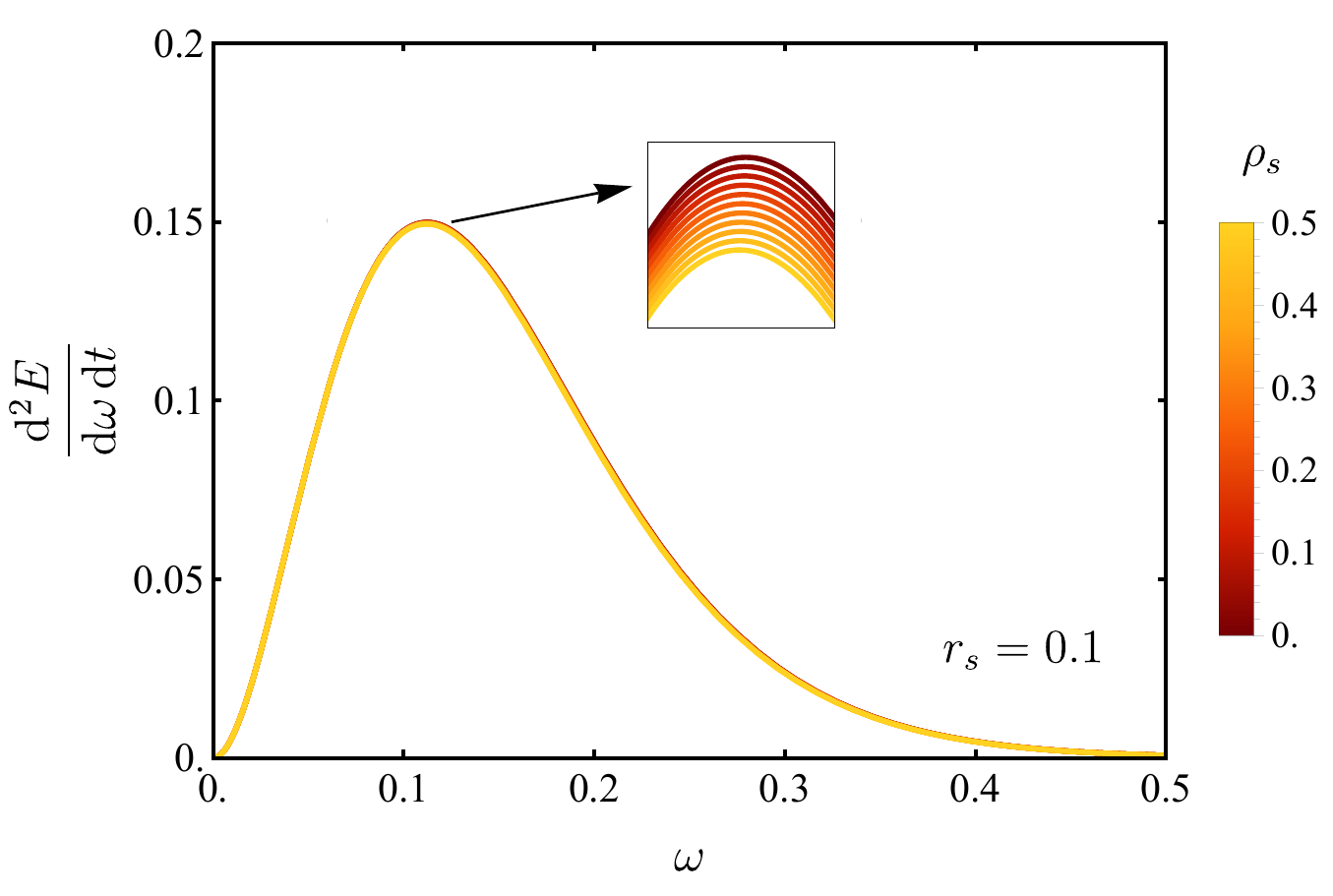} \hspace{2mm}
    \includegraphics[width=82mm]{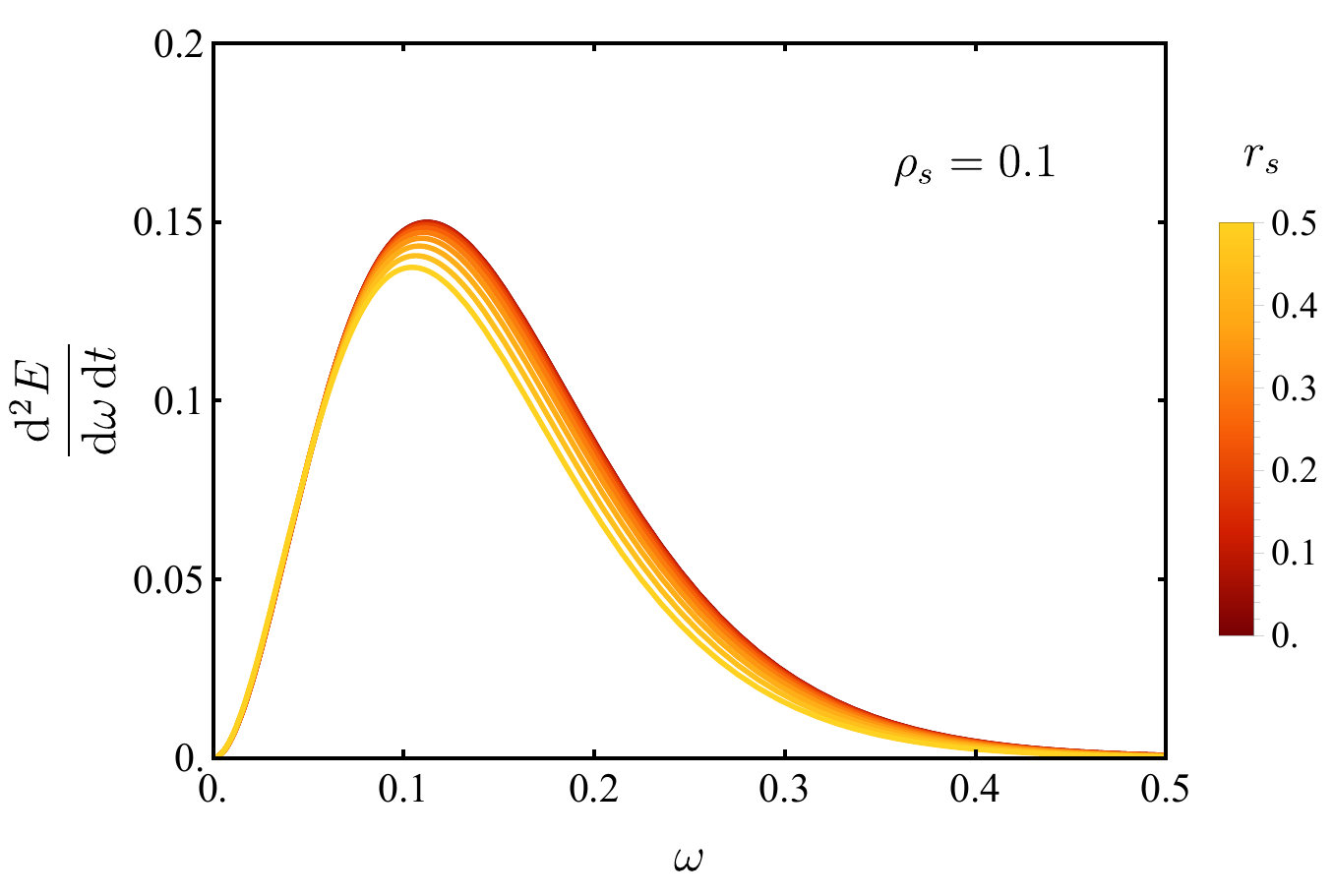}
    \caption{Energy flux as a function of the frequency $\omega$. The left panel shows configurations with $M=1$ and fixed $r_{\text{s}}=0.1$, while the right panel corresponds to fixed $\rho_{\text{s}}=0.1$ and varying $r_{\text{s}}$.}
    \label{fig:Emis}
\end{figure}

Attention is now turned to the properties of the energy flux radiated by the black hole
\ie
\label{energyemission}
	\frac{{{\mathrm{d}^2}E}}{{\mathrm{d}\omega \mathrm{d}t}} = \frac{{2{\pi ^2}{\sigma}_{lim}}}{{{e^{\frac{\omega }{T}}} - 1}} {\omega ^3}.
\fe
Figure~\ref{fig:Emis} shows the energy flux as a function of the frequency $\omega$. The left panel corresponds to configurations with $M=1$ and fixed $r_{\text{s}}=0.1$, while the right panel is obtained by keeping $\rho_{\text{s}}=0.1$ fixed and varying $r_{\text{s}}$. In agreement with the behavior found in the evaporation--time analysis, increasing either $\rho_{\text{s}}$ or $r_{\text{s}}$ results in a lower emission rate. 

We now turn to the associated particle production rate, which is given by
\ie
\frac{\mathrm{d}^{2}N}{\mathrm{d}\omega \mathrm{d}t}
= \frac{2\pi^{2}\,\sigma_{lim}\omega^{2}}
       {{{e^{\frac{\omega }{T}}} - 1}}.
\fe

\begin{figure}[ht]
    \centering
    \includegraphics[width=82mm]{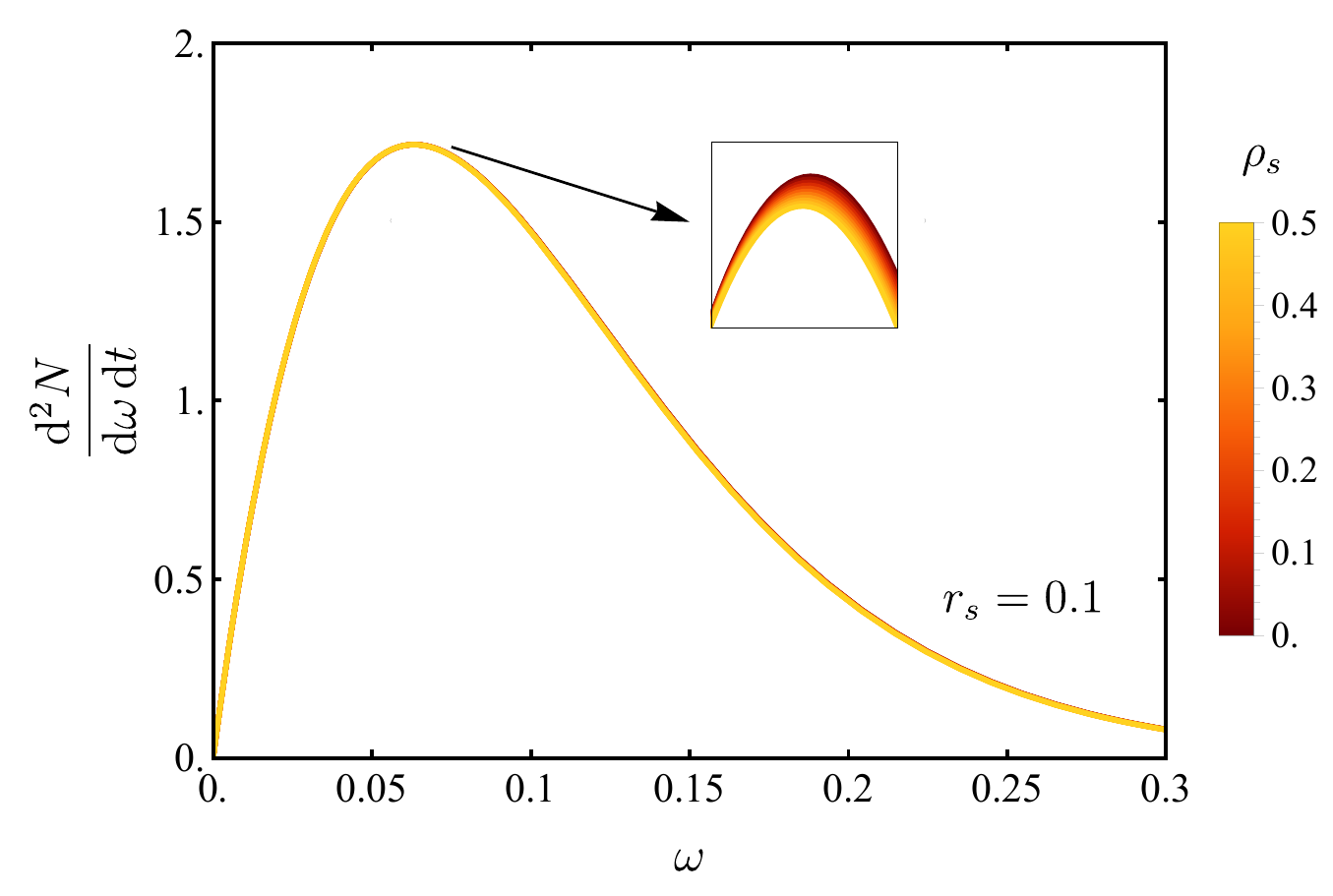} \hspace{2mm}
    \includegraphics[width=82mm]{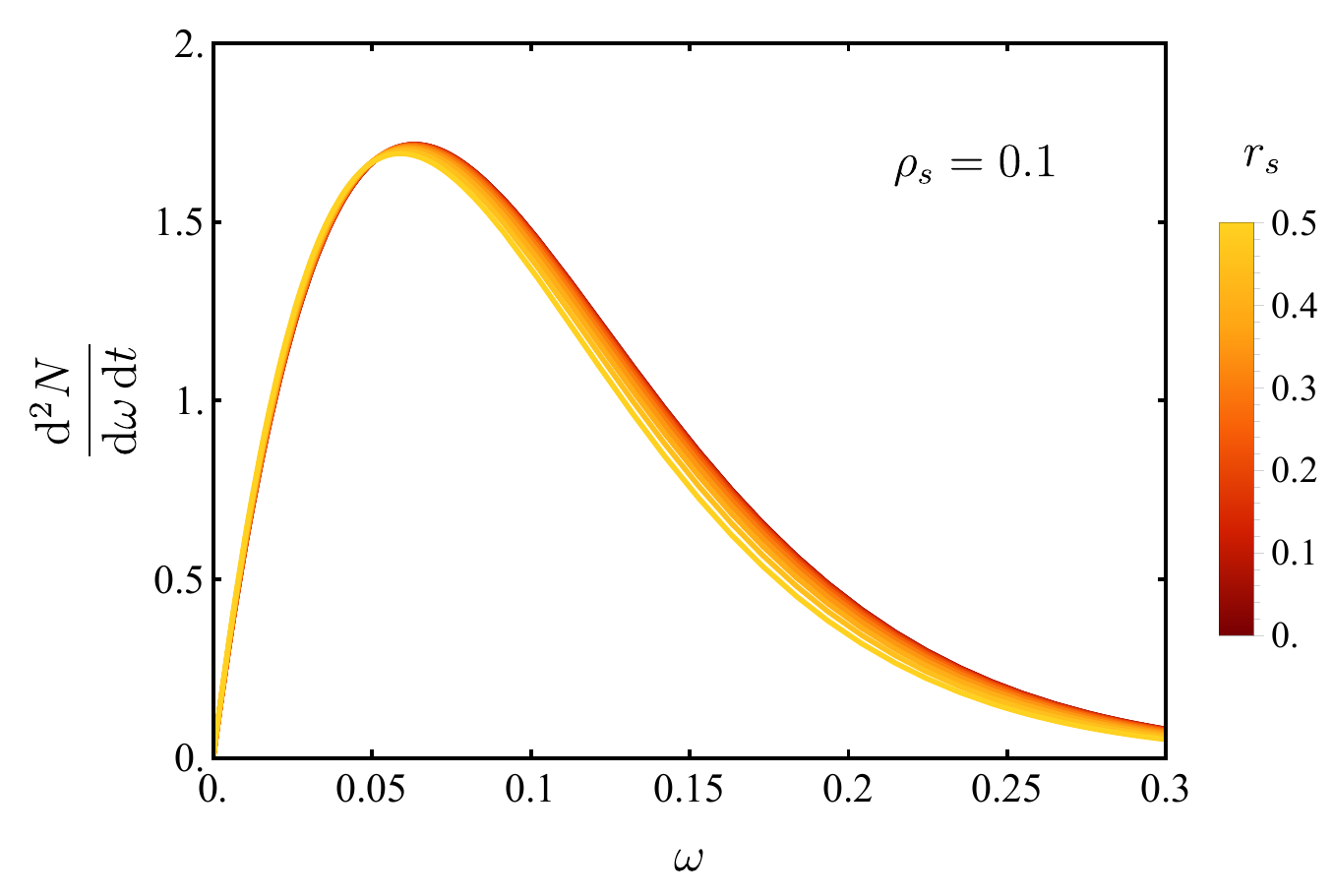}
    \caption{Particle emission rate as a function of the frequency $\omega$. The left panel corresponds to the case $M=1$ with $r_{\text{s}}=0.1$ held fixed, whereas the right panel is obtained by fixing $\rho_{\text{s}}=0.1$ and varying $r_{\text{s}}$.}
    \label{fig:EmiN}
\end{figure}

Figure~\ref{fig:EmiN} presents the particle emission rates for $M=1$, showing the particle flux as a function of the frequency $\omega$. The left panel corresponds to configurations with fixed $r_{\text{s}}=0.1$ and varying $\rho_{\text{s}}$, while the right panel is obtained by holding $\rho_{\text{s}}=0.1$ constant and varying $r_{\text{s}}$. In other words, increasing either $\rho_{\text{s}}$ or $r_{\text{s}}$ leads to a reduced the particle emission rate as well.

\section{Partial wave analysis}

In this section, we employ the partial–wave expansion to compute the absorption and scattering cross section. The massless scalar field is governed by 
\begin{equation}\label{eq:KG}
 \frac{1}{\sqrt{-g}} \partial_\mu \left(\sqrt{-g} \, g^{\mu\nu} \partial_\nu \Phi \right) = 0
\end{equation}

Starting from the metric Eq.~\eqref{eq:metric}, we introduce the usual separable ansatz for a scalar perturbation,
\begin{equation}
\Phi(t,r,\theta,\phi)=\frac{\psi(r)}{r}\,Y_{\ell m}(\theta,\phi)\,e^{-i\omega t},
\label{eq:separation}
\end{equation}
and substitute it into the wave Eq. \eqref{eq:KG} leads to the radial equation
\begin{equation}
\frac{\mathrm{d}}{\mathrm{d}r}\!\left[f(r)\frac{\mathrm{d}\psi}{dr}\right] 
+ \Big[\omega^{2}-V_{\text{eff}}(r)\Big]\psi=0,
\label{eq:radial}
\end{equation}
with effective potential 
\begin{equation}
V_{\text{eff}}=f(r)\!\left(\frac{\ell(\ell+1)}{r^{2}}
+\frac{f'(r)}{r}\right),
\label{eq:potential}
\end{equation}
where prime denotes the derivation with respect to the radial coordinate. By introducing the tortoise coordinate $r^*$ the Eq. \eqref{eq:radial} yields in a Schr\"{o}dinger--like equation
\begin{equation}
\frac{\mathrm{d}^2\psi}{\mathrm{d}{r^*}^2}+ \left(\omega^{2}-V_{\text{eff}}\right)\psi=0,
\label{eq:radial2}
\end{equation}
Taking into account Eq. \eqref{eq:potential}, the effective potential for the scalar field exhibits a localized barrier that vanishes asymptotically at the event horizon and at spatial infinity. The asymptotic form of the radial wavefunction $\psi(r^*)$ is therefore constrained: it must be purely ingoing at the horizon, while at infinity it consists of a superposition of incident and reflected waves. Imposing these physical boundary conditions yields ~\cite{macedo2015scattering,macedo2016absorption,anacleto2023absorption}
\begin{equation}\label{bound}
	\psi(r^*) \approx 
	\begin{cases}
	{\mathcal{R}_{\omega \ell}} \psi_{{I}} , &\text{for} \ {r^*} \to r_{\rm h} ~ (r \to -\infty )\\
	\psi_{{II}}+\mathcal{T}_{\omega \ell} \psi^*_{{II}},&\text{for} \ {r^*} \to \infty ~ (r \to \infty ).\\
	\end{cases}
\end{equation}
where $\psi_I$ and $\psi_{II}$ in Eq. \eqref{eq:radial2}, admit a expansion of the form \cite{macedo2013absorption} 
\begin{align}\label{Rroman1}
	{\psi_{{I}}}& = {e^{ - i\omega r*}}\sum\limits_{j = 0}^N {{{(r - {r_{\rm h}})}^j}a_{{r_{\rm h}}}^{j}}, \\ \label{Rroman2}
	{\psi_{{II}}}&= {e^{ -i\omega r*}}\sum\limits_{j = 0}^N {\frac{{a_\infty ^{j}}}{{{r^j}}}} .
\end{align}
To find the coefficients $a^j_{r_\text{h}}$ and $a^j_{\infty}$, we need to numerically solve the radial equation Eq. \eqref{eq:radial2} while imposing the boundary conditions given in Eq. \eqref{bound}. 
 The coefficients $\mathcal{R}_{\omega \ell}$ and $\mathcal{T}_{\omega \ell}$ corresponds to reflection and transmission coefficients via $|\mathcal{R}_{\omega \ell}|^2$ and $|\mathcal{T}_{\omega \ell}|^2$, respectively. Moreover the conserved relation of $|\mathcal{R}_{\omega \ell}|^2+|\mathcal{T}_{\omega \ell}|^2=1$ is satisfied.
The phase shifts are defined through the relation~\cite{dolan2009scattering}
\begin{equation}\label{phase}
	e^{2i\delta_{\omega \ell}} = (-1)^{l+1} \, \mathcal{R}_{\omega \ell}.
\end{equation}  
The phase shift enables us to extract the absorption and scattering cross sections.  
\section{Absorption cross section}

\begin{figure}[ht!]
    \centering
    \includegraphics[width=80mm]{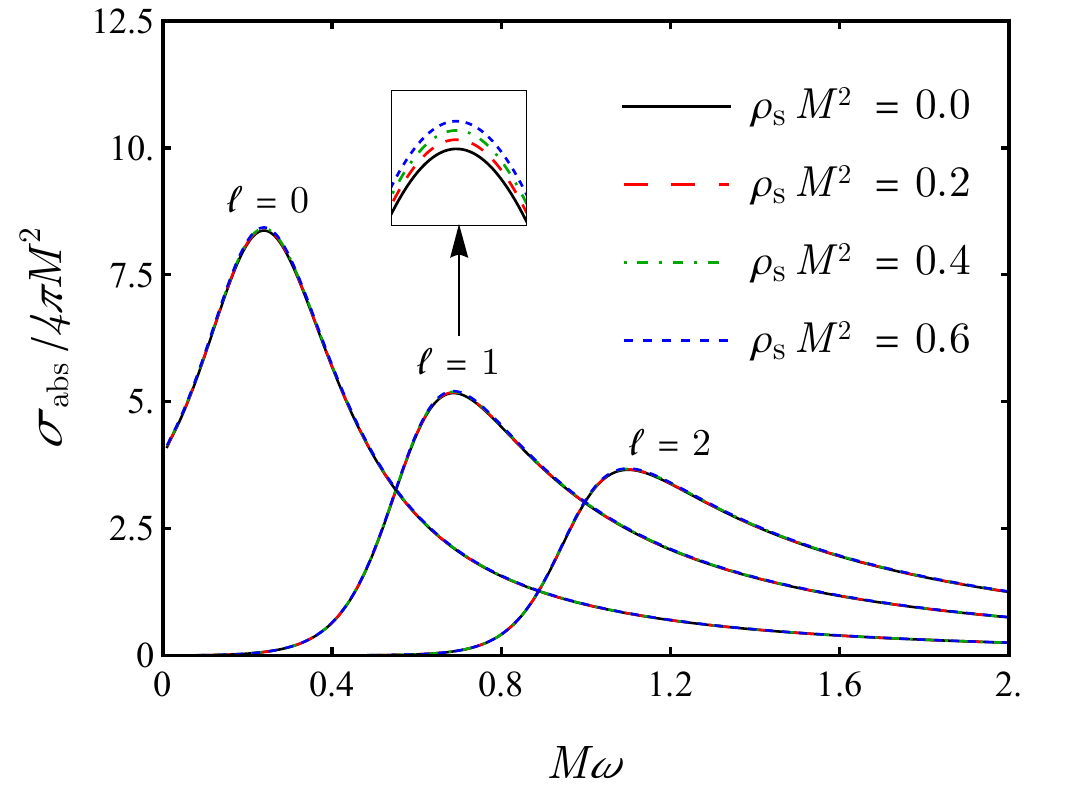} \hspace{2mm}
    \includegraphics[width=80mm]{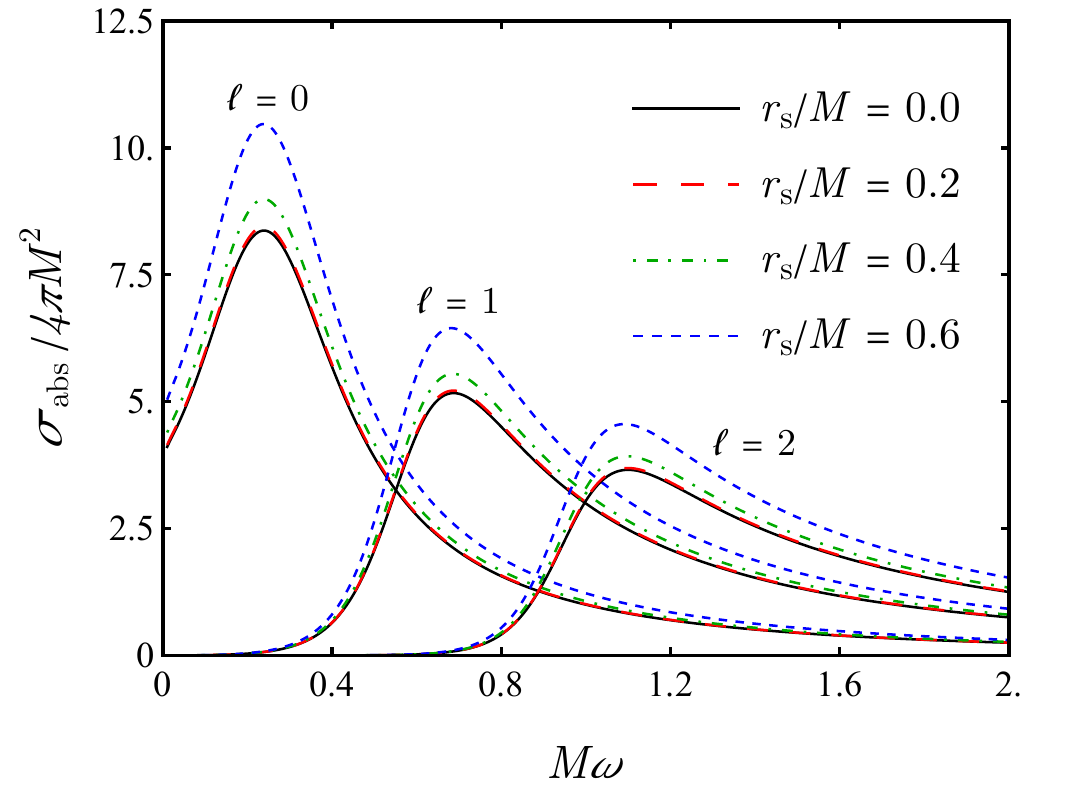} 
    \caption{ Partial absorption cross sections with respect to $M\omega$, for $\ell=0,1,2$. In the left panel, ${r_\text{s}}/M = 0.1$ and various Hernquist parameter ${\rho_\text{s}}M^2$ vary from $0.0$ to $0.6$. In the right panel, ${\rho_\text{s}}M^2$ is fixed at $0.1$ and different Hernquist parameter ${r_\text{s}}/M$ have been considered.}
    \label{fig:Psigma}
\end{figure}

Following the partial wave method \cite{futterman1986scattering}, the total absorption cross section is obtained as the sum of the individual partial contributions,
\begin{equation}
    \sigma_{\rm abs} = \sum_{\ell = 0}^{\infty} \sigma_{\rm abs}^{\ell},
\end{equation}
where the partial cross section is given by
\begin{equation}\label{partial1}
    \sigma_{\rm abs}^{\ell}
    = \frac{\pi}{\omega^{2}} (2\ell + 1)
      \bigl( 1 - |e^{2i\delta_{\omega \ell}}|^{2} \bigr).
\end{equation}
with \(\delta_{\omega \ell}\) denoting the phase--shift parameter defined in Eq. \eqref{phase}.
To compute the phase shifts associated, we employ the numerical procedure outlined in Ref.~\cite{macedo2016absorption,dolan2009scattering,heidari2024scattering}. In particular, we solve the radial wave equation in Eq.~\eqref{eq:radial2} and match the numerical solution to the analytic asymptotic forms~\eqref{Rroman1}--\eqref{Rroman2}. The integration is initialized close to the event horizon at 
\(
r \simeq r_\text{h}+10^{-3}r_\text{h} ,
\) 
and extended up to the asymptotically flat region at 
\(
r \simeq 100\, r_{\rm h},
\)
ensuring a controlled extraction of the large--radius behaviour. The series appearing in the asymptotic matching is truncated at the tenth order (\(N=10\)), which we have verified to be sufficient for numerical convergence within machine precision.
 By inserting the computed phase shift into Eq.~\eqref{partial1}, we evaluate the absorption cross section for multipole numbers $\ell=0,1,2$ and different values of \({r_\text{s}}/M\) and \({\rho_\text{s}} M^{2}\) in Fig.~\ref{fig:Psigma}, to assess the influence of the Hernquist dark--matter halo surrounding the black hole.

 In both the left and right panels of Fig.~\ref{fig:Psigma}, the characteristic structure of the absorption spectrum is evident: at low frequencies the monopole contribution (\(\ell = 0\)) dominates, while higher multipoles begin to contribute as \(\omega\) increases. In the left panel, the normalized Hernquist radius is fixed at \({r_\text{s}}/M = 0.1\), and the characteristic normalized density parameter \({\rho_\text{s}} M^2\) is varied between \(0\) and \(0.6\). The resulting curves indicate that the influence of \({\rho_\text{s}} M^2\) on the absorption cross section is generally very subtle. A magnified window around the \(\ell = 1\) mode is included to show this subtle effect more clearly: increasing \({\rho_\text{s}} M^2\) leads to a slight yet discernible enhancement of the corresponding partial absorption cross section.

The right panel instead explores the impact of varying \({r_\text{s}}/M\), while keeping \({\rho_\text{s}} M^2 = 0.1\) fixed. In this case, the dependence of the absorption spectrum is significantly more pronounced. For all multipole numbers displayed, the absorption cross section increases systematically with larger values of \({r_\text{s}}/M\). This trend becomes particularly pronouced near the peak of each curve, where the maxima shift upward as \({r_\text{s}}/M\) grows.

By comparing the variations associated with the normalized Hernquist parameters, we conclude that the scale radius \({r_\text{s}}/M\) produces a stronger effect on the absorption cross section than the density amplitude \({\rho_\text{s}} M^2\).

\begin{figure}[ht]
    \centering
    \includegraphics[width=82mm]{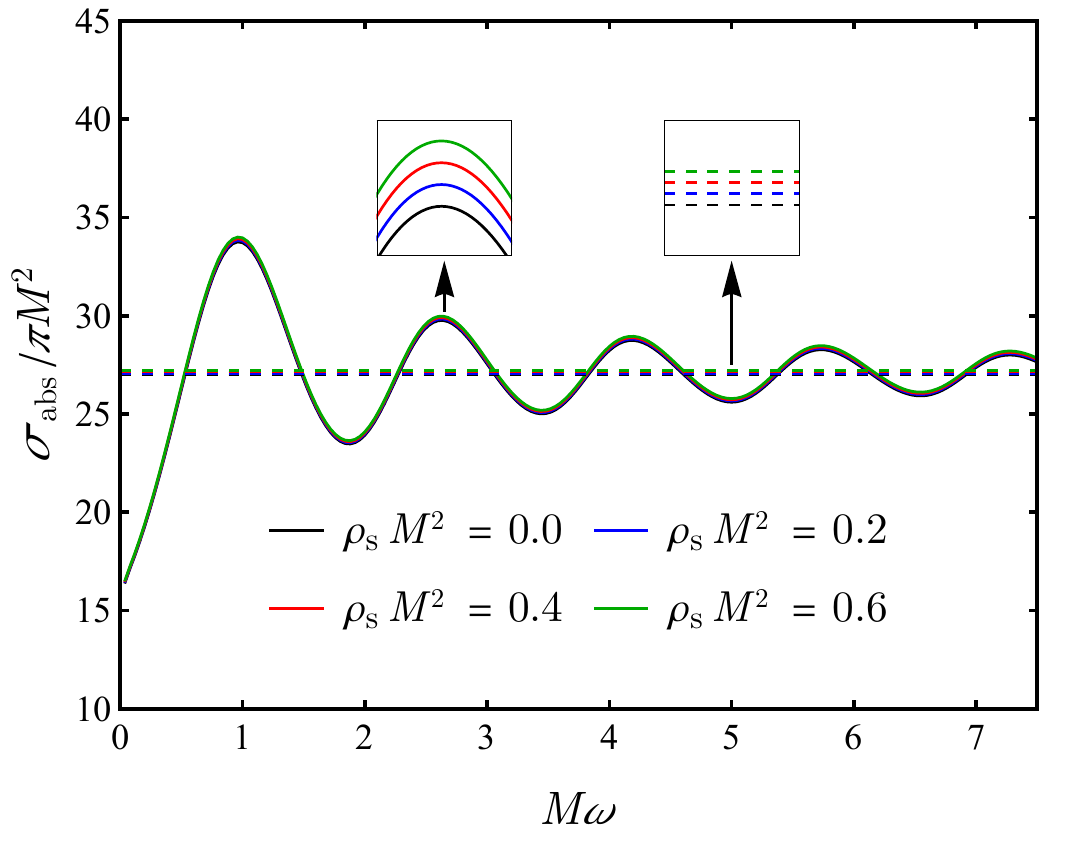} \hspace{2mm}
    \includegraphics[width=82mm]{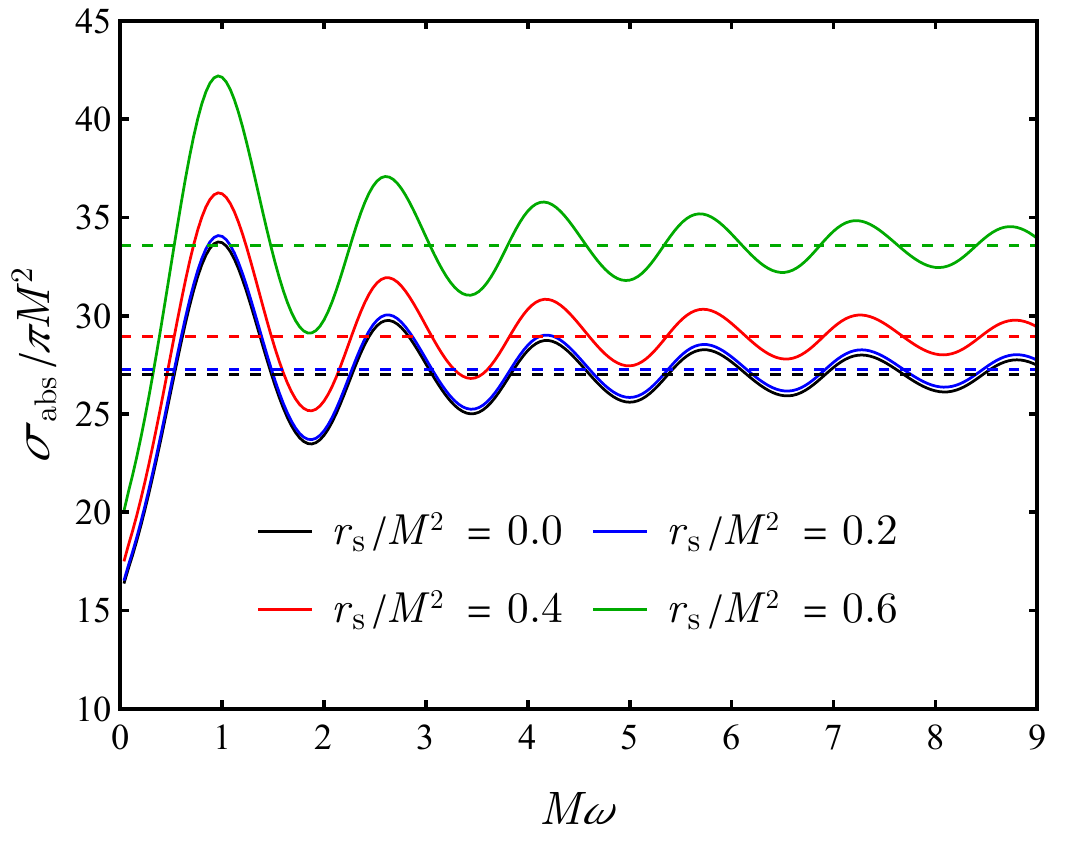}
    \caption{ Total absorption cross sections as a function of $M\omega$. 
In the left panel, the normalized scale radius is fixed at ${r_\text{s}}/M = 0.1$, while in the right panel, ${\rho_\text{s}} M^2$ is fixed at $0.1$. The dashed lines indicate the corresponding geometric absorption cross sections.}
    \label{fig:Tsigma}
\end{figure}

Moreover, Fig.~\ref{fig:Tsigma} shows the total absorption cross section, obtained by summing the first seven multipoles from $\ell = 0$ to $\ell = 6$. Consistent with what was observed for the partial modes in Fig.~\ref{fig:Psigma}, the influence of the density parameter ${\rho_\text{s}} M^2$ (left panel) is very subtle when the scale radius is fixed. For better illustration, a close inspection of the magnified region reveals a slight increase in the total cross section as ${\rho_\text{s}} M^2$ becomes larger. On the other hand, the effect of the scale radius ${r_\text{s}}/M$ (right panel) is clearly much stronger: increasing ${r_\text{s}}/M$ noticeably enhances the total absorption across the entire frequency range.

It is worth noting that the overall shape of the curves remains essentially unchanged; neither the position nor the pattern of the peaks in $M\omega$ is affected by variations of the Hernquist parameters. 

\subsection{Low--Frequency Regime}

At sufficiently low frequencies, the absorption cross--section of a black hole is known to converge to the area of its event horizon. This universal behavior has been extensively established in the literature and represents a robust feature of black hole scattering processes \cite{Das1996we,heidari2025absorption,Higuchi:2001si}. In the long--wavelength regime, where the incident wave probes length scales comparable to or larger than the black hole size, the detailed structure of the spacetime becomes subdominant. As a result, the absorption process is governed primarily by the geometric properties of the horizon, causing the cross--section to asymptotically approach the horizon area.

In particular, for scalar field perturbations propagating on a Schwarzschild background, the low--frequency absorption cross--section is analytically shown to coincide with the geometrical area of the event horizon. Remarkably, this limit persists across a wide range of scenarios, including rotating black holes, modified black hole parameters, and nontrivial surrounding geometries \cite{leite2017scalar,baptista2025scattering,macedo2013absorption}, provided the incident frequency remains sufficiently small.

Consequently, in the low--frequency regime, the absorption cross--section can be expressed as
\begin{align}\label{eq:lowfr}
\sigma_{\text{low}} &\simeq 4\pi r_{\text{h}}^{2} \\ \nonumber
&= \pi \left( 2M + 4\pi \rho {r_\text{s}}^{3} - {r_\text{s}} 
+ \sqrt{\left(2M + 4\pi \rho {r_\text{s}}^{3} - {r_\text{s}}\right)^{2} + 8M {r_\text{s}}} \right)^{2},
\end{align}
where $r_{\text{h}}$ denotes the radius of the event horizon.

\begin{figure}[ht]
    \centering
    \includegraphics[width=82mm]{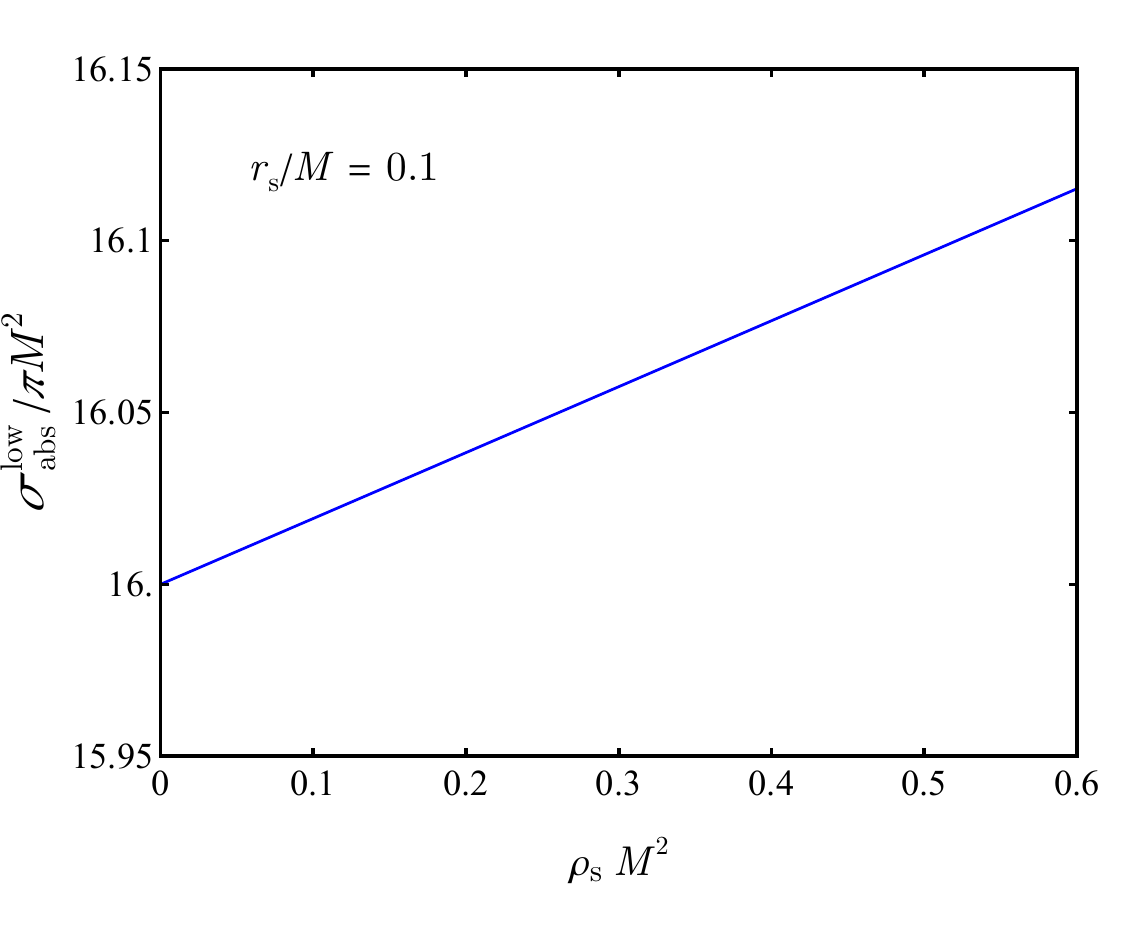} \hspace{2mm}
    \includegraphics[width=82mm]{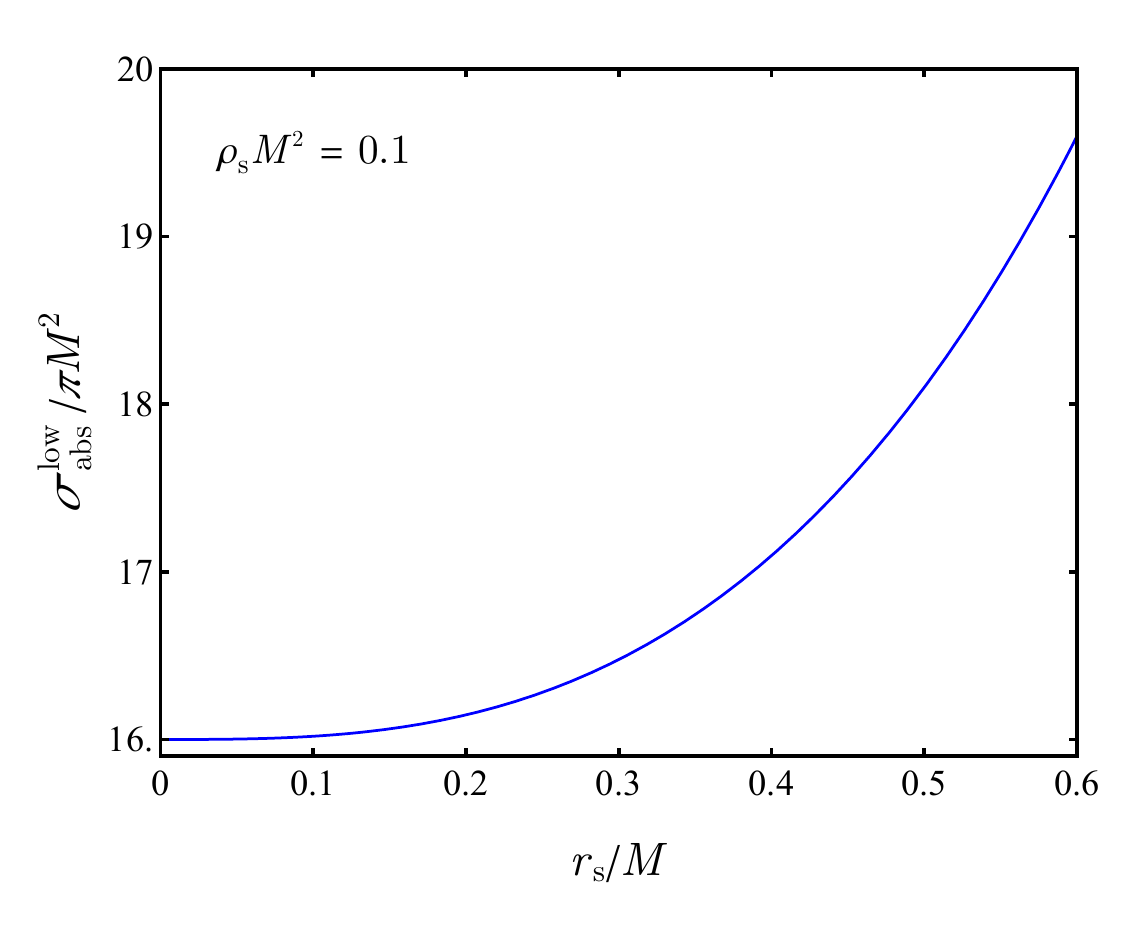}
    \caption{Absorption cross sections in the low-frequency regime as a function of the parameters ${\rho_\text{s}} M^2$ and $ {r_\text{s}}/M$ in the left and right panels, respectively. In the left panel, the normalized scale radius is fixed at ${r_\text{s}}/M = 0.1$, while in the right panel, ${\rho_\text{s}} M^2$ is fixed at $0.1$. }
    \label{fig:lowsigma}
\end{figure}
The low--frequency behavior of the absorption cross--section for different values of the Hernquist parameters is illustrated in Fig.~\ref{fig:lowsigma}. As expected, when both ${r_\text{s}}$ and ${\rho_\text{s}}$ vanish, the spacetime reduces to the Schwarzschild black hole, for which the low--frequency absorption cross--section approaches $\sigma_{\text{low}} \simeq 16\pi M^{2}$. For a fixed value of ${r_\text{s}}/M = 0.1$, increasing the dimensionless density parameter ${\rho_\text{s}} M^{2}$ from $0$ to $0.6$ produces only a negligible modification in the low--frequency absorption cross--section. This weak dependence is clearly visible in the left panel of Fig.~\ref{fig:Tsigma}, where $\sigma_{\text{abs}}$ remains close to the Schwarzschild case, $16\pi M^2$, at the low--frequency regime for different values of ${\rho_\text{s}} M^{2}$.

In contrast, the absorption cross--section exhibits a more pronounced sensitivity to variations in the scale radius ${r_\text{s}}$. As shown in the right panel of Fig.~\ref{fig:lowsigma}, increasing ${r_\text{s}}/M$ leads to a noticeable enhancement of the absorption cross--section. This trend is consistent with the behavior displayed in Fig.~\ref{fig:Tsigma}, where larger values of ${r_\text{s}}/M$ correspond to higher absorption cross--sections at low frequencies.


\subsection{High--Frequency Regime}

In this section, we investigate the high--frequency limit of the absorption cross--section, commonly referred to as the geometric cross--section, $\sigma_{\text{geo}}$. In this regime, wave effects become negligible and the absorption process is well described by null geodesic motion. The geometric cross--section is defined as
\begin{equation}\label{eq:geocross}
\sigma_{\text{geo}} = \pi b_c^2,
\end{equation}
where $b_c$ denotes the critical impact parameter separating scattered and captured null trajectories. To determine $b_c$, we analyze the dynamics of massless particles propagating in the black hole spacetime. The equations of motion for null geodesics follow from the Lagrangian
\begin{equation}
2\mathcal{L} = -f(r)\dot{t}^2 + f(r)^{-1}\dot{r}^2 + r^2\dot{\phi}^2 = 0,
\end{equation}
where an overdot denotes differentiation with respect to an affine parameter. Owing to the spacetime symmetries, the energy $E$ and angular momentum $L$ of the photon are conserved quantities. Introducing the impact parameter as $b = L/E$, the radial equation of motion can be written in the following form
\begin{equation}
\dot{r}^2 + \mathcal{V}_{\text{eff}} = 0,
\end{equation}
with the effective potential given by
\begin{equation}
\mathcal{V}_{\text{eff}} = E^2 - \frac{f(r)}{r^2}L^2.
\end{equation}

The critical impact parameter corresponds to unstable circular photon orbits, located at the photon sphere radius $r_{\text{ph}}$. This radius is obtained by imposing the conditions
\begin{equation}
\mathcal{V}_{\text{eff}} = \frac{\mathrm{d}\mathcal{V}_{\text{eff}}}{\mathrm{d}r} = 0,
\end{equation}
which lead to the following relations for the photon sphere and the critical impact parameter, respectively:
\begin{align}\label{cr}
&2f(r_{\text{ph}}) - r_{\text{ph}} f'(r_{\text{ph}}) = 0, \\
\label{impact}
&b_c = \frac{r_{\text{ph}}}{\sqrt{f(r_{\text{ph}})}}.
\end{align}

For the Hernquist black hole, solving Eq.~\eqref{cr} yields an explicit expression for the photonic radius as
\begin{align}\label{eq:rph}
r_{\text{ph}} & = \frac{1}{3}\Bigg(
 \sqrt[3]{\mathcal{D}}
+ \frac{
9M^2
+ 6M\!\left(6\pi {\rho_\text{s}} {r_\text{s}}^3 + {r_\text{s}}\right)
+ {r_\text{s}}^2 \left(1 - 6\pi {\rho_\text{s}} {r_\text{s}}^2\right)^2
}{
\sqrt[3]{\mathcal{D}}
}
+ 3M + 6\pi \rho_\text{s} {r_\text{s}}^3 - 2 {r_\text{s}}
\Bigg) \\
& \approx \, 3 M+ \frac{6 \pi  M \rho_{\text{s}} r_{\text{s}}^3 (9 M+2 r_{\text{s}})}{(3 M+r_{\text{s}})^2} \nonumber,
\end{align}
where the auxiliary quantity $\mathcal{D}$ is defined as
\begin{align}
&\mathcal{D} =
(3M + {r_\text{s}})^3
+ 9\pi {\rho_\text{s}} {r_\text{s}}^3 (18M^2 + {r_\text{s}}^2)
+ 108\pi^2 {\rho_\text{s}}^2 {r_\text{s}}^6 (3M - {r_\text{s}})
+ 216\pi^3 {\rho_\text{s}}^3 {r_\text{s}}^9 \nonumber \\
&+ \left(
27\pi {\rho_\text{s}} {r_\text{s}}^5 \Big[
2(3M + {r_\text{s}})^3
+ \pi {\rho_\text{s}} {r_\text{s}}^3 (180M^2 - 96M {r_\text{s}} - 25 {r_\text{s}}^2)
+ 8\pi^2 {\rho_\text{s}}^2 {r_\text{s}}^6 (9M + 13 {r_\text{s}})
- 144\pi^3 {\rho_\text{s}}^3 {r_\text{s}}^9
\Big]
\right)^{1/2},
\end{align}
and the second line of Eq. (\ref{eq:rph}) correspond to the approximation of $\rho_{\text{s}}$ small.

Substituting Eq.~\eqref{eq:rph} into Eq.~\eqref{impact}, the critical impact parameter can be obtained, which in turn allows us to compute the high--frequency absorption cross--section. The final expression for $\sigma_{\text{geo}}$ reads
\begin{align}\label{sigmahigh}
\sigma_{\text{geo}} =
\frac{\pi
\Bigl(
3M - 2{r_\text{s}} + 6\pi {\rho_\text{s}} {r_\text{s}}^3
+ \dfrac{\mathcal{B}}{\sqrt[3]{\mathcal{D}}}
+ \sqrt[3]{\mathcal{D}}
\Bigr)^2}
{9\Biggl[
1
- \dfrac{6M}{3M - 2{r_\text{s}} + 6\pi {\rho_\text{s}} {r_\text{s}}^3
+ \dfrac{\mathcal{B}}{\sqrt[3]{\mathcal{D}}}
+ \sqrt[3]{\mathcal{D}}}
- \dfrac{12\pi {\rho_\text{s}} {r_\text{s}}^3}{3M + {r_\text{s}} + 6\pi {\rho_\text{s}} {r_\text{s}}^3
+ \dfrac{\mathcal{B}}{\sqrt[3]{\mathcal{D}}}
+ \sqrt[3]{\mathcal{D}}}
\Biggr]},
\end{align}
where the parameter $\mathcal{B}$ is given by
\begin{equation}
\mathcal{B} =
9M^2
+ {r_\text{s}}^2 \left(1 - 6\pi {\rho_\text{s}} {r_\text{s}}^2\right)^2
+ 6M \left({r_\text{s}} + 6\pi {\rho_\text{s}} {r_\text{s}}^3\right).
\end{equation}

Figure~\ref{fig:Tsigma} displays the total absorption cross--section obtained from partial waves with $l = 0$ to $l = 6$ for different values of the Hernquist parameters. The corresponding geometric cross--section, computed from Eq.~\eqref{sigmahigh}, is shown by dashed horizontal lines. As expected, the total absorption cross--section approaches a constant value in the high--frequency regime, confirming its convergence toward the geometric cross--section $\sigma_{\text{geo}}$. 

Moreover, the high--frequency absorption cross--section exhibits a stronger sensitivity to variations in the scale parameter ${r_\text{s}}/M$ than to changes in the density parameter ${\rho_\text{s}}M^2$. As shown in the left panel of Fig.~\ref{fig:Tsigma}, the modification of the geometric cross--section induced by varying ${\rho_\text{s}}M^2$ is extremely weak and remains practically indistinguishable. In contrast, the right panel demonstrates that $\sigma_{\text{geo}}$ increases with increasing ${r_\text{s}}/M$, indicating that the geometric scale of the system plays a dominant role in determining the high--frequency absorption behavior. 

\section{Scattering cross section}

The scattering of waves by a black hole can be systematically described using a partial wave expansion
\cite{futterman1986scattering,dolan2013scattering,dolan2009scattering,anacleto2020absorption,leite2019black,baptista2025scattering}.
Within this framework, the scattering amplitude is expressed as
\begin{equation}
g(\theta) =
\frac{1}{2i\omega}
\sum_{\ell=0}^{\infty}
(2\ell+1)\left( e^{2i\delta_{\ell}} - 1 \right)
P_{\ell}(\cos\theta),
\label{eq:amplitude}
\end{equation}
where $P_\ell(\cos\theta)$ denotes the Legendre polynomial of degree $\ell$, and $\delta_\ell$ is the phase shift associated with the $\ell$th partial wave, as defined in Eq.~\eqref{phase}. The entire angular dependence of the scattering process is encoded in the phase shifts through this expansion.

The observable angular distribution of the scattered radiation is characterized by the differential scattering cross--section, which is related to the scattering amplitude by \cite{sanchez1978elastic}
\begin{equation}
\frac{\mathrm{d}\sigma}{\mathrm{d}\Omega} = \left| g(\theta) \right|^2.
\label{eq:diff_cross_section}
\end{equation}
This quantity provides direct information about the angular structure of the scattering process.

The total scattering cross--section is obtained by integrating the differential cross--section over the full solid angle. For a monochromatic wave of frequency $\omega$, this leads to
\begin{equation}
\sigma_{\text{sca}} =
\int \frac{\mathrm{d}\sigma}{\mathrm{d}\Omega}\, \mathrm{d}\Omega
= \frac{\pi}{\omega^2}
\sum_{\ell=0}^{\infty}
(2\ell+1)
\left| e^{2i\delta_\ell} - 1 \right|^2.
\label{eq:total_cross_section}
\end{equation}
This expression highlights that the total scattering cross--section arises from a coherent superposition of all partial waves, with each contribution weighted by the corresponding phase shift.  

\subsection{Partial scattering cross--section}

\begin{figure}[ht!]
    \centering
    \includegraphics[width=80mm]{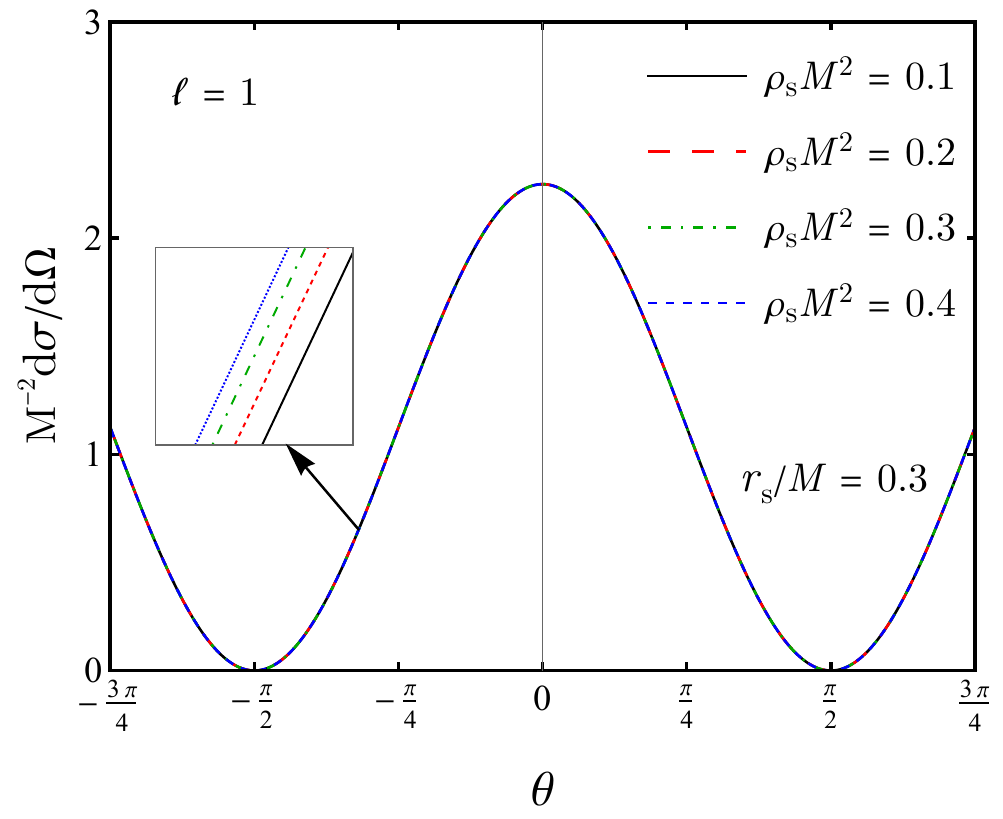}
    \includegraphics[width=80mm]{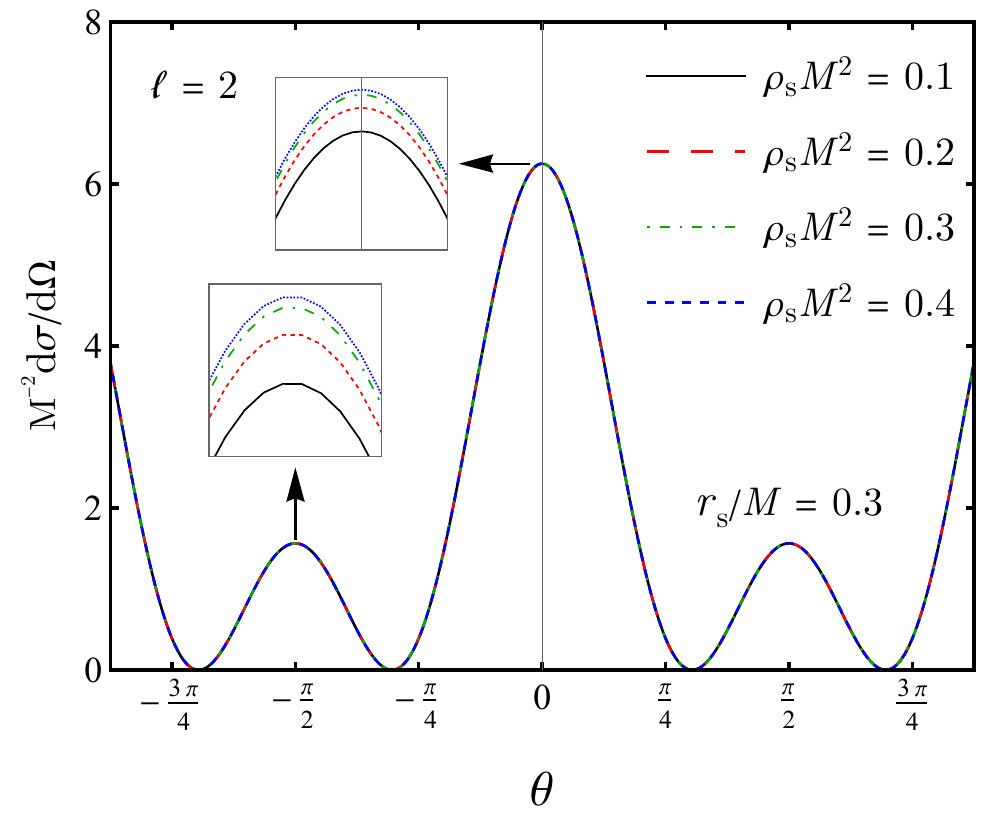}
    \caption{  The normalized partial differential scattering cross--sections of the scalar wave for $M\omega = 1$, $\ell = 1,~2$, ${r_\text{s}}/M = 0.3$ and various different Hernquist parameter ${\rho_\text{s}}M^2$.}
    \label{fig:Sct01}
\end{figure}

\begin{figure}[ht!]
    \centering
    \includegraphics[width=80mm]{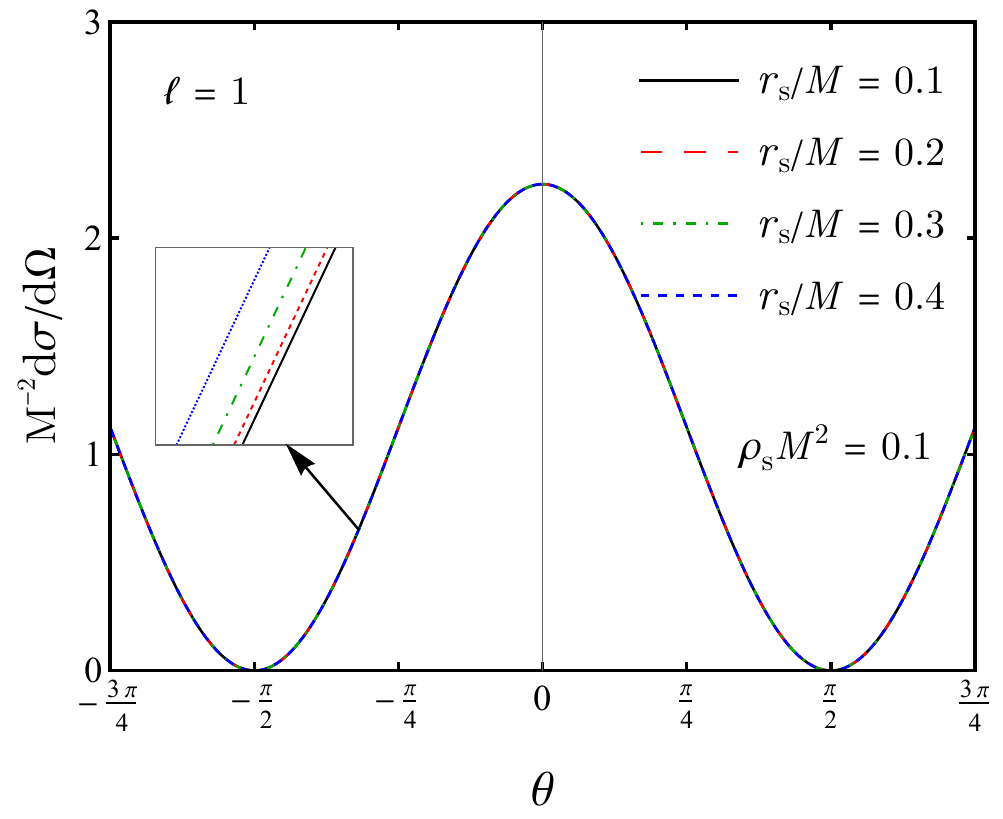}
    \includegraphics[width=80mm]{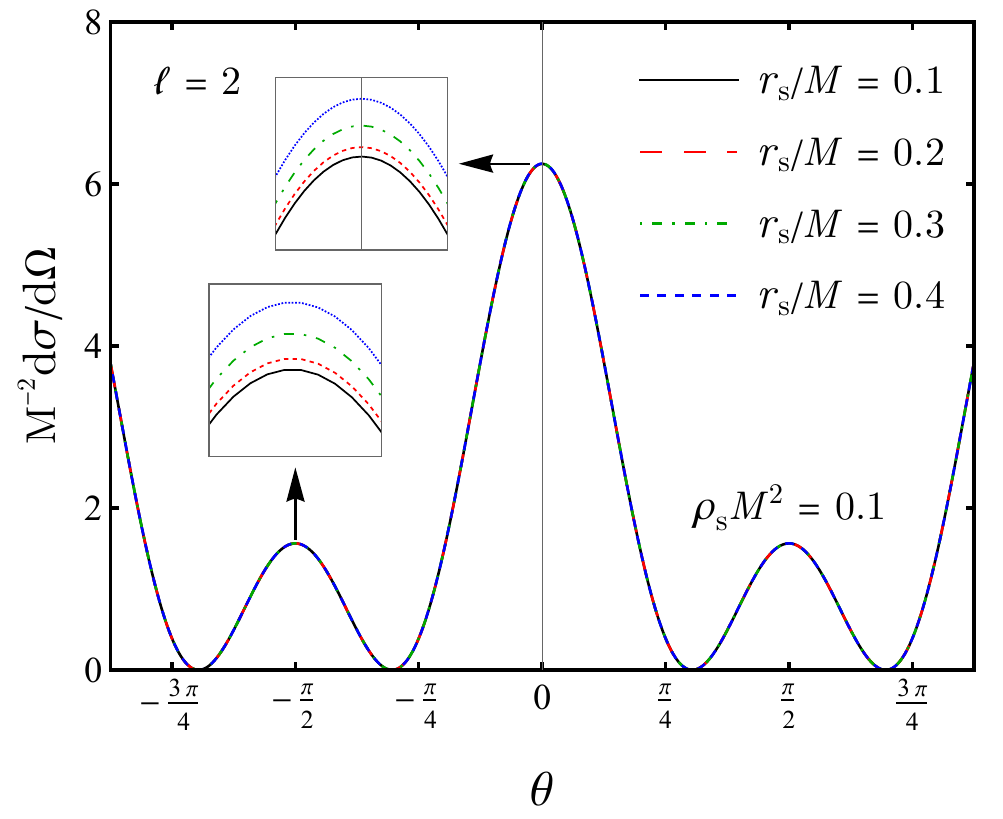}
    \caption{  The normalized partial differential scattering cross--sections of the scalar wave for $M\omega = 1$, $\ell = 1,~2$, ${\rho_\text{s}}M^2 = 0.1$ and various different Hernquist parameter ${r_\text{s}}/M$.}
    \label{fig:Sct02}
\end{figure}

To investigate the influence of the Hernquist black hole parameters on the scattering properties, we analyze the angular dependence of the differential scattering cross--section for selected multipole numbers. In particular, Figs.~\ref{fig:Sct01}--\ref{fig:Sct02} display the differential cross--section as a function of the scattering angle $\theta$ for several representative values of $\ell$, illustrating how individual partial waves contribute to the overall scattering pattern.

Figure~\ref{fig:Sct01} presents the partial differential scattering cross-sections for scalar waves with $M\omega = 1$ and ${r_\text{s}}/M = 0.3$, for several choices of ${\rho_\text{s}} M^{2}$, comparing the cases $\ell = 1$ and $\ell = 2$, in the left and right panle, respectively. In both multipole numbers, a clear and consistent pattern emerges: increasing ${\rho_\text{s}} M^{2}$ leads to a higher overall amplitude of the partial scattering cross-section. While the $\ell = 1$ and $\ell = 2$ curves display their usual differences in angular sharpness, their dependence on ${\rho_\text{s}} M^{2}$ remains uniform, indicating that the effect of this parameter is broadly insensitive to the multipole structure.
Figure~\ref{fig:Sct02} displays the partial differential scattering cross-sections for scalar waves with fixed parameters $M\omega = 1$ and ${\rho_\text{s}} M^{2} = 0.1$, considering the multipole modes $\ell = 1$ and $\ell = 2$ in the left and right panels, respectively. For both modes, the qualitative behavior is the same: the partial scattering cross-section increases as the Hernquist parameter ${r_\text{s}}/M$ becomes larger. Although the $\ell = 1$ and $\ell = 2$ curves differ in the shape of their angular patterns, they exhibit the same monotonic dependence on ${r_\text{s}}/M$.
\begin{figure}[ht!]
    \centering
    \includegraphics[width=80mm]{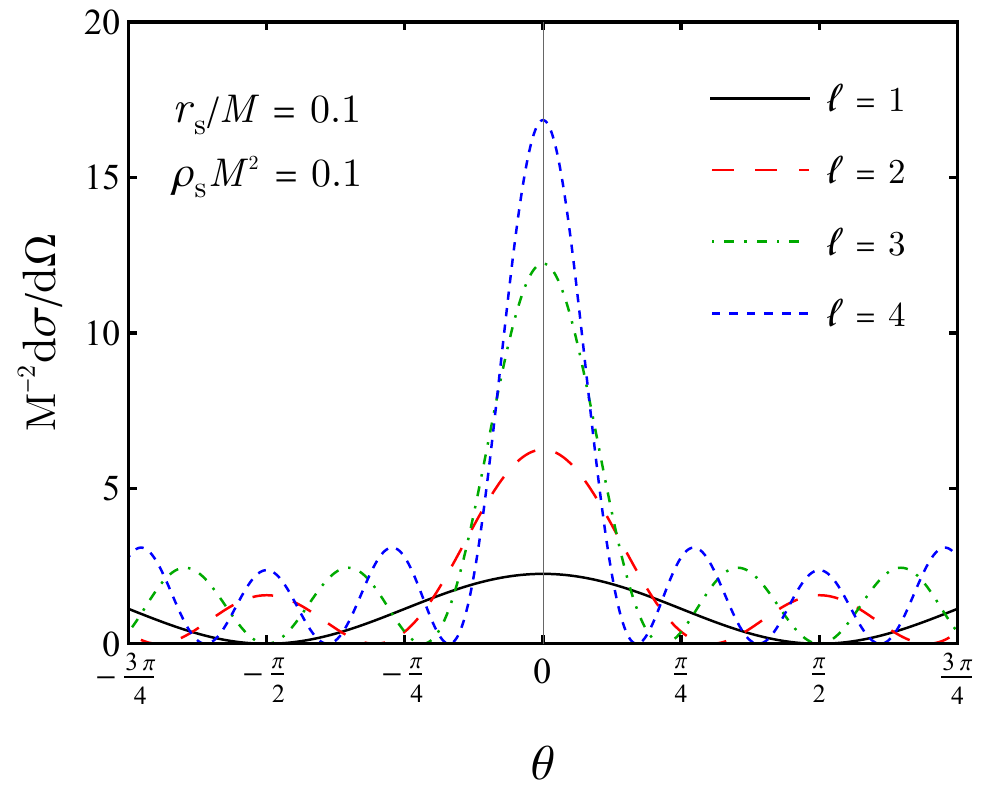}
    \caption{  The normalized partial scattering cross--sections for different multipole numbers when ${\rho_\text{s}} M^2={r_\text{s}}/M=0.1$.}
    \label{fig:Sct03}
\end{figure}
Figure~\ref{fig:Sct03} displays the scalar wave differential scattering cross-sections in the Hernquist black hole spacetime for several multipole numbers, evaluated at ${\rho_\text{s}} M^2 = {r_\text{s}}/M = 0.1$. As the angular momentum mode $\ell$ increases, from $\ell = 1$ to $4$, the angular structure of the scattering becomes progressively sharper and more oscillatory, with the dominant contribution concentrated near the forward direction where the cross-section reaches its maximum. This behavior is consistent with the expected enhancement of higher multipole components, reflecting the increased angular resolution of the partial waves.  
\subsection{Total scattering cross section}

Numerical evaluation of the scattering amplitude via the partial-wave expansion in Eq. \eqref{eq:amplitude} exhibits poor convergence due to the oscillatory nature of the Legendre polynomials. To obtain accurate numerical results, we employ a convergence acceleration method, originally developed by Yennie et al. \cite{yennie1954phase} and later adapted to black hole scattering contexts \cite{dolan2006fermion}. This iterative regularization procedure systematically redefines the series coefficients, effectively suppressing high-$\ell$ oscillations and yielding a rapidly convergent representation. We applied second--order regularized series, convergence is achieved for the analyzed frequency range by including partial waves up to $\ell \thicksim 50$.
\begin{figure}[ht!]
	\centering
	\includegraphics[width=80mm]{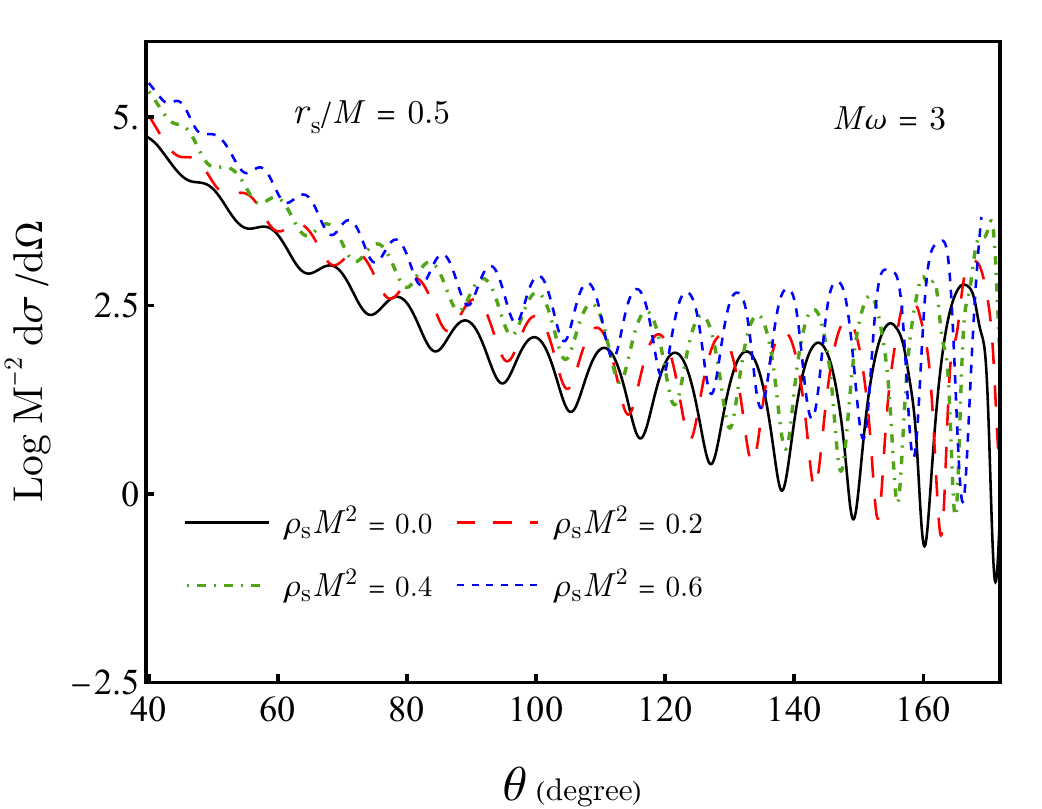}   \includegraphics[width=80mm]{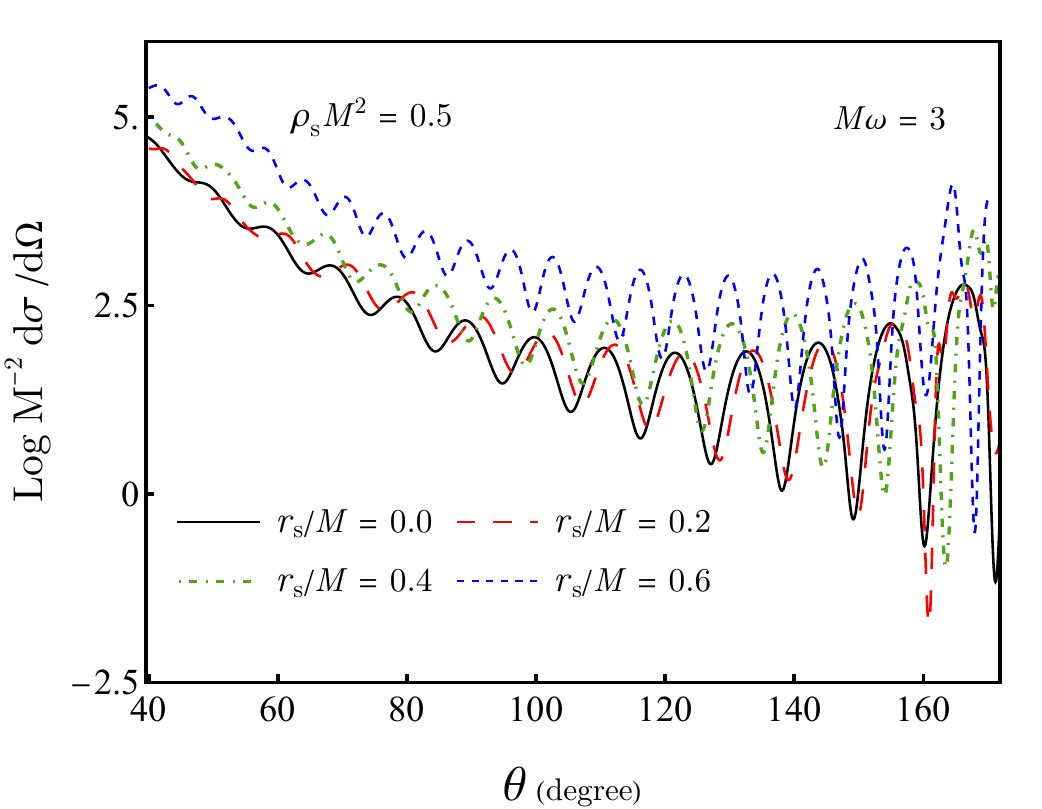} 
	\caption{ Total scattering cross--section in logarithmic scale for a Hernquist black hole at fixed frequency $M\omega = 3$.
Left panel: Dependence on the density parameter ${\rho_\text{s}} M^2$, varied from $0$ to $0.6$, with the scaled radius fixed at ${r_\text{s}}/M = 0.5$.
Right panel: Dependence on the scaled radius ${r_\text{s}}/M$, varied from $0$ to $0.6$, with the density parameter fixed at ${\rho_\text{s}} M^2 = 0.5$.
}
	\label{fig:Tscat}
\end{figure}
The angular dependence of the total scattering cross--section, displayed in logarithmic scale as a function of the scattering angle $\theta$, is shown in Fig.~\ref{fig:Tscat}. 

In the left panel, where the scaled radius is fixed at ${r_\text{s}}/M = 0.5$, increasing the central density parameter ${\rho_\text{s}} M^2$ leads to a moderate enhancement of the forward scattering peak accompanied by a slight narrowing of the fringe pattern. In contrast, the right panel demonstrates that variations in the scaled radius ${r_\text{s}}/M$, for a fixed density ${\rho_\text{s}} M^2 = 0.5$, produce a more pronounced effect. In this case, the total scattering cross--section increases significantly, and the interference fringes become markedly narrower as ${r_\text{s}}/M$ increases from $0$ to $0.6$.

These results indicate that the scattering properties of the Hernquist black hole are more sensitive to changes in the scale radius ${r_\text{s}}$ than to variations in the density parameter ${\rho_\text{s}}$.  

\section{Geodesics}

In gravitational physics, geodesic motion provides the fundamental link between spacetime geometry and the dynamics of free particles. The trajectories of such particles encode information about curvature and symmetries inherent in the metric. In systems where a Hernquist dark matter halo surrounds a central black hole, this analysis takes on particular significance: the halo contributes an extended mass distribution that modifies the spacetime geometry in a non-trivial way. Studying the motion of test particles—both massive and massless—thus offers a clear probe into how the combined gravitational field, shaped by both the central compact object and the dark matter environment, alters particle paths.  In general, the motion of a free particle in a curved spacetime is governed by the geodesic equation reads
\begin{equation}
\frac{\mathrm{d}^{2}x^{i}}{\mathrm{d}{\lambda}^{2}} + \Gamma^i_{j k} \frac{\mathrm{d}x^{j}}{\mathrm{d}{\lambda}}\frac{\mathrm{d}x^{k}}{\mathrm{d}{\lambda}} = \epsilon.
\label{Eq:Geo}
\end{equation}

where $\Gamma^i_{j k}$ are the 
Christoffel symbols, 
$\lambda$ is an affine parameter, 
and where $\epsilon$ equals to $0$ 
for massless particles and $-1$ for massive particles. The equations yield four relations for each spacetime coordinate These four equations for Hernquist black hole are as follows
\begin{align}\label{geot}
&\frac{\mathrm{d}^2 t}{\mathrm{d}{\lambda}^2} +
\frac{r' t' \left(\frac{2 M}{r^2}+\frac{4 \pi  \rho_{\text{s}}  r_{\text{s}}^3}{(r+r_{\text{s}})^2}\right)}{1-\frac{2 M}{r}-\frac{4 \pi  \rho_{\text{s}}  r_{\text{s}}^3}{r+r_{\text{s}}}}=\epsilon,\\ \label{geor}
&\frac{\mathrm{d}^2r}{\mathrm{d}{\lambda}^2}  +\frac{1}{(r+r_{\text{s}})^3}\Big[\frac{(r+r_{\text{s}})^2 r'^2 \left(M (r+r_{\text{s}})^2+2 \pi  \rho_{\text{s}}  r^2 r_{\text{s}}^3\right)}{r \left((r-2 M) (r+r_{\text{s}})-4 \pi  \rho_{\text{s}}  r r_{\text{s}}^3\right)}\\  \nonumber
&\frac{\left((2 M-r) (r+r_{\text{s}})+4 \pi  \rho_{\text{s}}  r r_{\text{s}}^3\right) \left(t'^2 \left(M (r+r_{\text{s}})^2+2 \pi  \rho_{\text{s}}  r^2 r_{\text{s}}^3\right)-r^3 (r+r_{\text{s}})^2 \left(\theta '^2+\sin ^2\theta \varphi '^2\right)\right)}{r^3}\Big]=\epsilon,\\ \label{geotheta}
&\frac{\mathrm{d}\theta^{\prime}}{\mathrm{d}{\lambda}} -\sin \theta  \cos \theta  \varphi '^2-\frac{2 \theta ' r'}{r}=\epsilon,\\ \label{geophi}
&\frac{\mathrm{d}\varphi^{\prime}}{\mathrm{d}{\lambda}} +\frac{2 \varphi ' \left(r'+r \theta ' \cot \theta \right)}{r}=\epsilon.
\end{align}
Here, prime denotes derivative taken with respect to the affine parameter. To explore the trajectory of a particle (massless or massive), Eq. \eqref{geot} - Eq. \eqref{geophi} must be integrated simultaneously.

\subsection{Light--like Geodesics}
\begin{figure}[ht!]
	\centering
	\includegraphics[width=80mm]{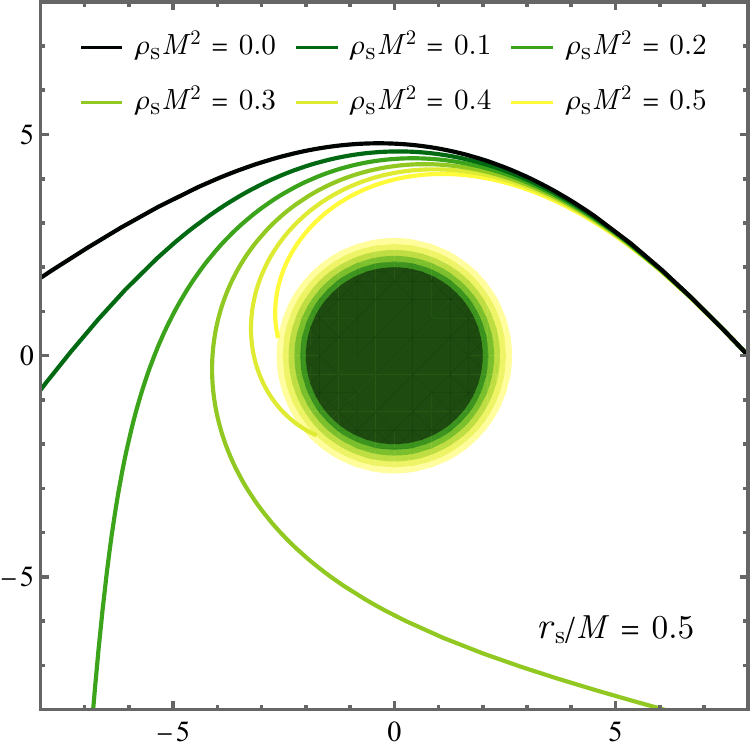}   
    \includegraphics[width=80mm]{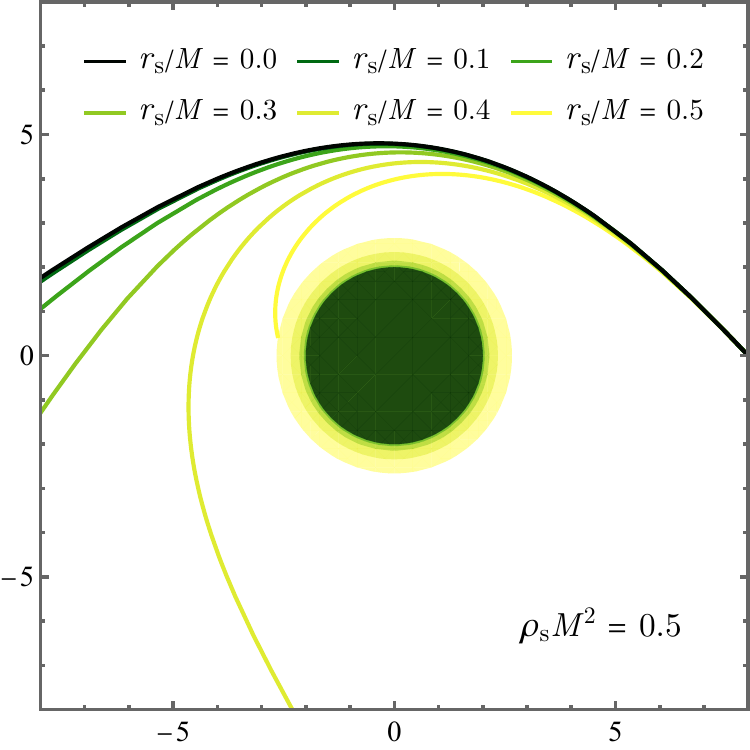} 
	\caption{Photon trajectories around a black hole within a Hernquist dark matter halo for variation of density parameter while scale radius in the mass unit is fixed at $0.5$.}
	\label{fig:GeoNull}
\end{figure}

Fig. \ref{fig:GeoNull} illustrates the influence of the Hernquist dark matter parameters on null geodesics. The left panel demonstrates the evolution of photon trajectories as the scale radius \(r_s/M\) varies from $0$ to $0.5$, while the dimensionless coupling parameter \( \rho_{\mathrm{s}} M^2\) is fixed at $0.5$. Conversely, the right panel shows the effect of varying \(\\rho_{\mathrm{s}} M^2\) over the same range while keeping \(r_{\mathrm{s}} = 0.5\) constant. In both cases, an increase in either parameter (\(r_{\mathrm{s}}/M\) or \(\rho_{\mathrm{s}} M^2\)) enhances the gravitational bending of light, drawing the photon orbits closer to the central black hole. 

A particularly noteworthy observation is the differential sensitivity of the lensing to these parameters. The photon paths exhibit a stronger response to increases in the coupling parameter \(\rho_\text{s}M^2\) compared to equivalent changes in \(r_{\mathrm{s}}/M\) for the chosen parameters. This is manifest in the right panel, where higher values of \(\rho_\text{s}M^2\) lead to a significantly stronger gravitational lensing effect, causing photons to be captured by the black hole's event horizon more swiftly. This behavior indicates that both dark matter density and the halo's scale play an important role in modifying the local curvature surrounding the black hole; however, the scale radius has a weaker impact.

\subsection{Time--like Geodesics}

\begin{figure}[ht!]
	\centering
	\includegraphics[width=80mm]{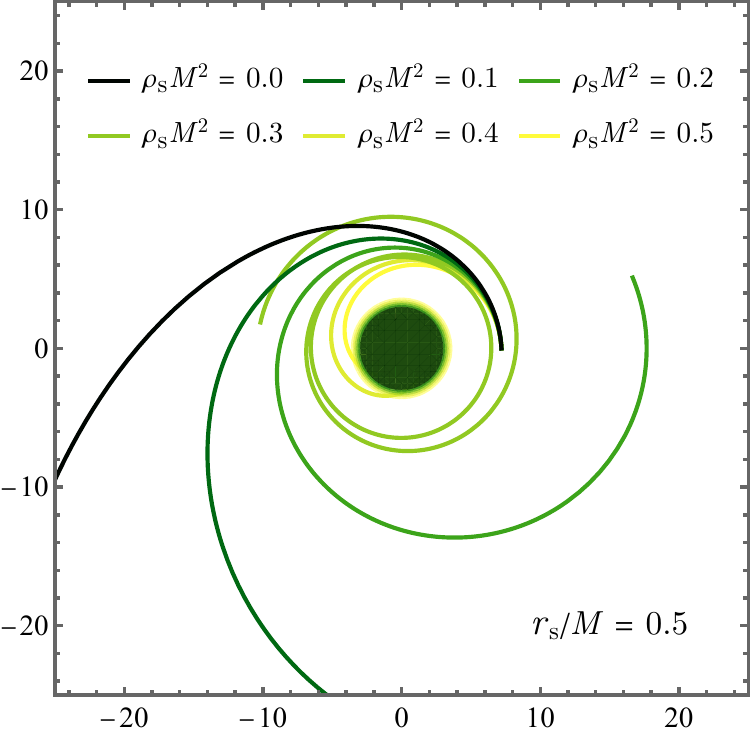}   \includegraphics[width=80mm]{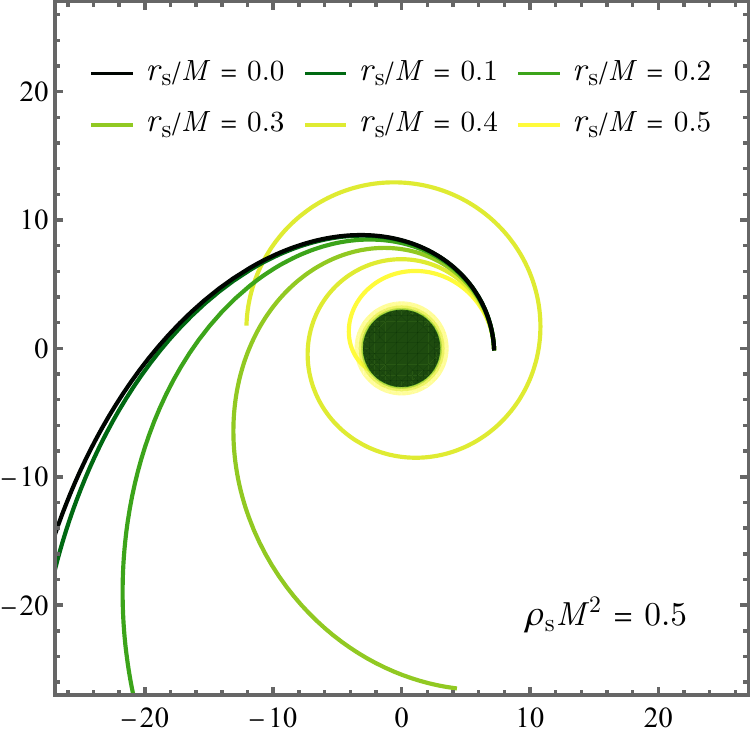} 
	\caption{Massive particle trajectories in the spacetime of a black hole embedded in a Hernquist dark matter halo for different scale radii when the density is set to $0.5$.}
	\label{fig:GeoMass}
\end{figure}

The behavior of massive particle trajectories can be another tool to explore the influence of the Hernquist dark matter halo on the spacetime geometry.  The result of the numerical solution of Eq. \eqref{geot} - Eq. \eqref{geophi} for a massive particle is calculated by considering $\epsilon = -1$. Fig.~\ref{fig:GeoMass} corresponding to the results presents the trajectories of a massive particle with different parameters. 

The left panel of Fig.~\ref{fig:GeoMass} illustrates the evolution of a particle as the scale radius \(r_{\mathrm{s}}/M\) varies, with the density parameter fixed at \(\rho_{\mathrm{s}} M^2 = 0.5\). The right panel shows trajectories for varying \(\rho_{\mathrm{s}} M^2\) while \(r_{\mathrm{s}}/M\) is held constant at $0.5$. A direct visual comparison along rays of the same initial conditions reveals a systematic increase in gravitational bending with increasing values of either parameter. 

However, the system displays a higher sensitivity to the density parameter \(\rho_{\mathrm{s}}M^2\) than to the scale radius \(r_{\mathrm{s}}/M\). For instance, a configuration with \(\rho_{\mathrm{s}}M^2 = 4\) and \(r_{\mathrm{s}}/M = 0.5\) (left panel), the particle is gravitationally trapped into the black hole. In stark contrast, for an equivalent configuration where \(r_{\mathrm{s}}/M = 0.4\) and \(\rho_{\mathrm{s}}M^2 = 0.5\) (right panel), while experiencing stronger deflection, the particle ultimately escapes capture. 
This result underscores that the density parameter modifies the local curvature more profoundly.



\section{Conclusion}

In this work we examined quantum and classical processes associated with a Schwarzschild black hole embedded in a Hernquist dark matter halo. Using the exact static and spherically symmetric solution describing this configuration, we investigated particle production, evaporation, absorption, scattering, and geodesic motion, with particular emphasis on how the halo parameters $\rho_{\text{s}}$ and $r_{\text{s}}$ modify the standard Schwarzschild behavior.

Quantum particle creation was analyzed for both scalar and fermionic fields. Hawking radiation was derived independently through Bogoliubov transformations and through the tunneling formalism with energy conservation. Both approaches led to the same effective temperature, while explicitly showing that increasing the dark matter density suppresses the occupation numbers for bosons and fermions. The tunneling description further reveals deviations from a strictly thermal spectrum once backreaction is taken into account, with the emission probability depending explicitly on the emitted frequency.

The evaporation process was studied in the high–frequency regime, where the absorption cross section approaches its geometric limit and greybody factors become negligible. In this approximation, the presence of the Hernquist halo reduces the energy and particle emission rates and increases the total evaporation time relative to the vacuum Schwarzschild case. The late–time evolution is governed by a remnant mass determined by the halo parameters, which constrains the physically admissible range of $\rho_{\text{s}}$ and ensures positivity of both the temperature and the remnant mass.

We also investigated wave propagation in this background through a partial–wave analysis of massless scalar perturbations. The phase shifts, partial absorption cross sections, and total absorption and scattering cross sections were computed numerically. While variations in the density parameter $\rho_{\text{s}}$ lead to relatively mild changes, the scale radius $r_{\text{s}}$ produces a more pronounced enhancement of the absorption cross section across different multipole modes, particularly near the peaks of the spectra.

Finally, null and timelike geodesics were analyzed to characterize light propagation and particle motion in the presence of the dark matter halo. The Hernquist profile modifies the effective potential governing both photon trajectories and massive particle orbits, altering deflection properties and orbital structure when compared with the Schwarzschild geometry.

As a possible extension, it would be natural to investigate quantum information aspects of the present black hole, including entanglement degradation and the associated HBAR entropy, along lines similar to Ref.~\cite{AraujoFilho:2025nmc}. In addition, exploring neutrino oscillations in this background appears to be a relevant direction, given the sensitivity of flavor evolution to spacetime modifications, as discussed in Refs.~\cite{Shi:2025plr,Shi:2025rfq}.

\section*{Acknowledgments}
\hspace{0.5cm}

N. H. is grateful for the support provided by three COST Actions: CA21106 (COSMIC WISPers in the Dark Universe: Theory, Astrophysics and Experiments), CA21136 (Addressing Observational Tensions in Cosmology with Systematics and Fundamental Physics, also known as CosmoVerse), and CA23130 (Bridging High and Low Energies in Search of Quantum Gravity, or BridgeQG). Also, N. H. is supported by Conselho Nacional de Desenvolvimento Cient\'{\i}fico e Tecnol\'{o}gico — CNPq, project number 152891/2025-0. In addition, A. A. Araújo Filho is supported by Conselho Nacional de Desenvolvimento Cient\'{\i}fico e Tecnol\'{o}gico (CNPq) and Fundação de Apoio à Pesquisa do Estado da Paraíba (FAPESQ), project numbers 150223/2025-0 and 1951/2025. Additionally, we thank Professor Caio F. B. Macedo and Professor S. Dolan for the fruitful discussions on the scattering cross section.


	\bibliography{main}
	\bibliographystyle{unsrt}
	
\end{document}